\documentclass[truedoublespace,tocchapterhead]{ccw_chithesis}
\usepackage{graphicx,float}
\usepackage{psfrag}
\usepackage{amssymb}
\usepackage[bf, tight, normalsize, FIGTOPCAP]{subfigure}

\newcommand{\bg}{{\mathbf{\nabla}}}
\newcommand{\p}{{\partial}}

\newcommand{\beq}{\begin{equation}}
\newcommand{\eeq}{\end{equation}}
\newcommand{\bqa}{\begin{eqnarray}}
\newcommand{\eqa}{\end{eqnarray}}

\newcommand{\bu}{{\mathbf u}}

\newcommand{\rd}{{\rm d}}
\renewcommand{\t}{{\tau}}

\begin{document}

\title{Azimuthal Asymmetries and Vibrational Modes in Bubble Pinch-off}
\author{Laura E. Schmidt}
\department{Physics}
\division{Physical Sciences}
\degree{Doctor of Philosophy}
\date{December 2008}
\maketitle

\begin{abstract}
The pressure-driven inertial collapse of a cylindrical void in an
inviscid liquid is an integrable, Hamiltonian system that forms a
finite-time singularity as the radius of the void collapses to zero.
Here it is shown that when the natural cylindrical symmetry of the void
is perturbed azimuthally, the perturbation modes neither grow nor decay,
but instead cause constant amplitude vibrations about the leading-order
symmetric collapse. Though the amplitudes are frozen in time, they grow relative
to the mean radius which is collapsing to zero, eventually overtaking the
leading-order symmetric implosion.  Including weak viscous dissipation destroys
the integrability of the underlying symmetric implosion, and the effect on the
stability spectrum is that short-wavelength disturbances are now erased as the
implosion proceeds.
Introducing a weak rotational flow component to the symmetric implosion dynamics
causes the vibrating shapes to
spin as the mean radius collapses.
The above theoretical scenario is compared to a closely related
experimental realization of void implosion: the disconnection of an air bubble from
an underwater nozzle.  There, the thin neck connecting the bubble
to the nozzle implodes primarily radially inward and disconnects.
Recent experiments were able to induce vibrations of the neck shape
by releasing the bubble from a slot-shaped nozzle.
The frequency and amplitude of the observed vibrations are consistent
with the theoretical prediction once surface tension effects are taken into account.
\end{abstract}

\acknowledgments{I am especially grateful to Wendy Zhang and Sidney Nagel for their constant
guidance, advice, generosity with their time, and for teaching
me to think about the `big picture'.  I will always cherish the inspiring times we spent
discussing physics in your offices, or on the lawn, or over cookies and coffee.
I also thank Leo Kadanoff and  Mark Oreglia for their support and scientific critique during
the completion of this thesis.
Special thanks are owed to Nathan Keim for sharing his
experimental data and images for this thesis, but also for sharing his ideas
in our many conversations throughout this collaboration.
During my time at the U. of Chicago
I encountered people who enhanced my own understanding of this work -- they are
Justin Burton, Margo Levine, Lipeng Lai, Robert Schroll, and Konstantin Turitsyn.

When I needed a break or encouragement, my family was always there.
For all of the support and trust over the years, thank you.
}
\tableofcontents
\listoffigures

\mainmatter

\chapter{Introduction}
One of the most fundamental questions in fluid mechanics,
is how does a volume of fluid break into two?
It happens all the time, all around us -- for example a stream
of water falling from a faucet breaks into droplets, then
these droplets splash as they hit the sink, sending off
even more tiny satellite droplets.
Despite the ubiquitous nature of the phenomena,
we still do not have a complete understanding of the
process for a generic combination of fluids~\cite{has94,egg97}.
The question is even more interesting because when a fluid
undergoes a topological transition, a singularity occurs in the
governing Navier-Stokes equations, where physical quantities
such as the liquid pressure or velocity diverge in a finite
time~\cite{pumir92,barenblatt96,leo97}.
The mathematical appeal of singularity formation
and technological relevance of the generation
of droplets and bubbles has spurred much interest in
topological changes of fluid surfaces in past years.
Many of the theoretical works have concentrated on characterizing
the universal features of the approach to the singularity~\cite{day98, leppinen03, egg07}.

\begin{figure}
\includegraphics[width=0.7\textwidth, angle=0]{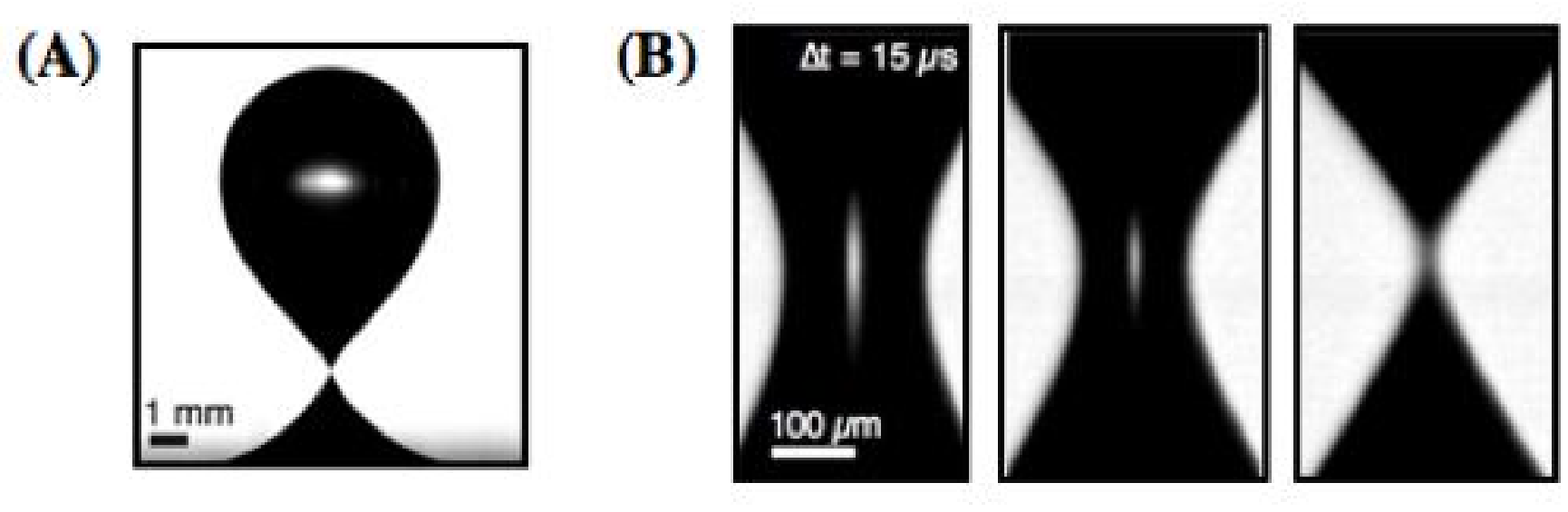}
\includegraphics[width=0.2\textwidth, angle=0]{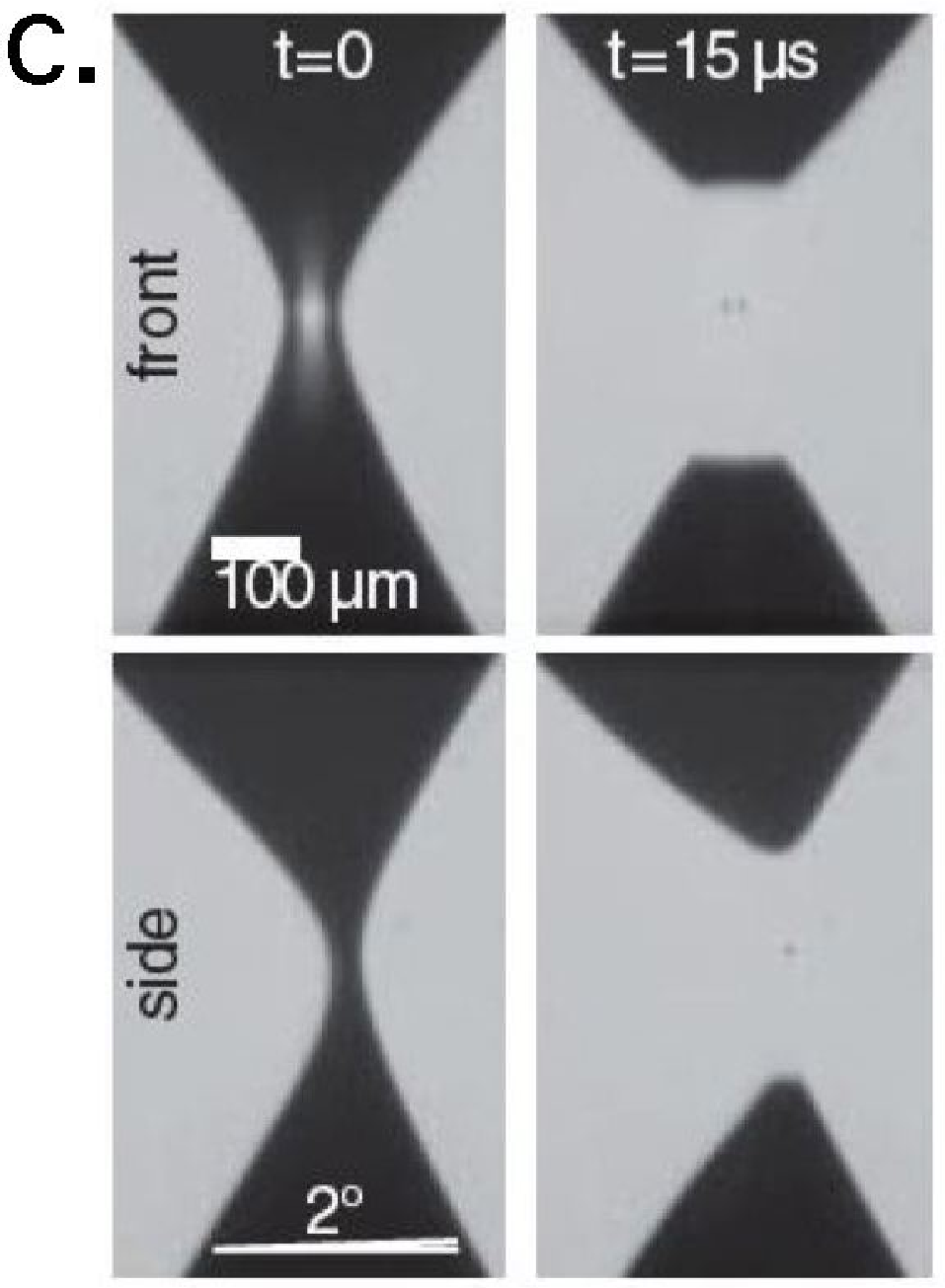}
\caption{A. Air bubble released from underwater nozzle, images from
N. Keim (U. Chicago) B. Close-up
images of radial implosion of neck as it disconnects.  C. Nozzle is
tilted by two degrees and shapes before and after disconnection show
signs of the asymmetry.  Two tiny side-by-side satellite bubbles are
left behind, evidence of the inherent memory in this system.
Nozzle is tilted towards viewer in the `front' view, and
to the left side in the `side' view.}
\label{bubbles}
\end{figure}

However, recent results from multiple experimental groups have
indicated that the pressure-driven implosion of a gas cavity
in an inviscid liquid is \textit{not} universal and that small
asymmetries present in the initial state are preserved. The retention
of such disturbances has
the potential to dramatically alter the final state at breakup.
In these experiments, a cavity is created and then disconnected either by
quasi-statically releasing a bubble from a nozzle submerged
in a liquid (see Fig.~\ref{bubbles})
~\cite{burton05, keim06, thoroddsen07, duclaux07, schmidt08},
by impacting a solid disc into a liquid bath~\cite{bergmann06, gekle08, bergmann08},
or by placing a bubble in a co-flowing stream~\cite{gordillo07}.
The cavity forms following the bubble as it rises,
following the disc as it passes into the bath, or in the middle
of the bubble as it is pulled apart by the divergent flow.
For all cases, the final dynamics before disconnection is
dominated by a rapid inertia-driven implosion, not by surface
tension which drives the breakup of a liquid drop in air~\cite{day98, shi94, chen97}.
In the only detailed investigation of the effect of initial azimuthal
asymmetry on the bubble pinch-off dynamics Keim \textit{et al} ~\cite{keim06} found
that tilting the nozzle by $2^{\circ}$ results in side-by-side
satellite bubble formation and non-circular neck cross sections.  Tilts
of even less than a tenth of a degree result in visibly asymmetric neck profiles,
demonstrating that the generic form of pinch-off realized in experiments
is asymmetric. Despite the fact that azimuthal asymmetry is usually
observed in the experiments, it has only recently begun to be explored
theoretically~\cite{schmidt08}.

The inertial implosion of a gas cavity belongs to a different
class of dynamics which are not universal, but instead have a precise
memory of the details of the initial state.  To understand the
origin of the observed memory, consider a theoretically simpler
but closely related system -- the implosion of a cylindrical void
in an inviscid liquid.
Integrable systems such as this have a perfect memory (with as many conserved
quantities as independent variables), and when they are perturbed,
the Kolmogorov-Arnold-Moser theorem predicts that the dynamics
will have trajectories which, for the most part, closely follow
those of the original integrable dynamics~\cite{ott}.
In this system,
which is both integrable and singular,
the underlying memory manifests itself in a particularly simple way --
as the amplitudes of shape disturbances become frozen in time as the cavity
implodes~\cite{schmidt08}.
The constant-amplitude disturbances cause vibrations of the void's
cross section shape as the implosion proceeds, with frequencies proportional
to $1/\tau$, where $\tau$ is the time until the radius closes to zero. Since the
frequency diverges as $\tau\rightarrow 0$, the shapes chirp, vibrating faster
and faster as the void closes.
This presents a natural explanation why the experiments observe
a variety of disconnection scenarios depending on the specific experimental
conditions.  Since the experiments have a dynamics close to an
integrable dynamics, small perturbations are retained as the
hole closes.

The first purpose of this paper
is to investigate additional physical effects which are mathematically
interesting and experimentally relevant.  Including weak viscous dissipation
in the model destroys the integrability of the leading order implosion
dynamics, and including a non-zero circulation to the liquid flow prevents the formation
of a singularity.  Both effects have consequences on how a disturbance
on the void surface evolves over time,
but within the experimentally accessible regime and for
large wavelength (small mode number) disturbances do not qualitatively
change the freezing and chirping behavior.
The second purpose is to directly connect the theoretical model
to the experimental realization for both the symmetric and perturbed cases.
This requires including surface tension, which, though it becomes
asymptotically irrelevant compared to inertial forces, has lingering effects
on the vibration frequency within the experimental regime.

We begin by reviewing the cylindrically symmetric
implosion dynamics and demonstrate that the dynamics can be
recast into Hamiltonian form.  We examine the model's
implications for quasi-2D (nearly cylindrical) void collapse and
show that though the integrable dynamics is fundamentally \textit{non}-universal,
it appears self-similar in this limit.
Then we consider the stability of the cylindrically symmetric
implosion when subjected to azimuthal shape perturbations and
derive the mode amplitude freezing and chirping behavior,
including asymptotic logarithmic corrections.
Next, we investigate changes to the leading-order symmetric dynamics
and vibrational mode behavior after taking into account viscous dissipation
and introducing a rotational flow.
Finally, we discuss the relation between experiments
for the cylindrically symmetric and azimuthally perturbed cases,
ending with a direct comparison between the theoretical prediction for
a single vibrational mode and an experimentally induced vibration
in the neck of a bubble released from an oblong nozzle.

\chapter{Axisymmetric cavity implosion}
\label{axi}
The implosion of an arbitrarily shaped cavity is a very complicated,
fully 3D problem.
However, with additional geometric constraints the problem is simplified.
The dynamics of a spherical cavity have been studied for over a century
\cite{besant59, rayleigh17, landaufluid}, and is an integral component in the
understanding cavitation~\cite{plesset77} and the phenomena of
sonoluminescence~\cite{hilgenfeldt96}.
The inertial collapse of a cylindrical cavity was first studied in relation
to bubble production from underwater nozzles~\cite{longuet91,oguz93} and
has more recently been applied to cavity collapse in liquids after solid body
impact~\cite{bergmann06} and cavities in granular media~\cite{lohse04}.
The inner gas dynamics can also play a role in the asymptotic
dynamics~\cite{gordillo05}. Here, we briefly reexamine the cylindrically
symmetric implosion and its implications so that we can later characterize its
stability when perturbed azimuthally.

The problem under consideration is the inertial implosion of an
infinite cylindrical vacuum cavity in an inviscid liquid.
After the void is created it begins
to implode inwards because the pressure inside the void is less than the
pressure of the liquid outside.
The inviscid fluid flow is governed by Euler equations:
$\p \vec{u}/\p t+ \vec{u}\cdot \vec{\nabla}\vec{u}=-\vec{\nabla} P/\rho$),
where $\vec{u}$ is the local velocity field, $P$ the pressure, and $\rho$ the
liquid density.
When incompressible, the continuity equation for volume conservation
$\vec{\nabla} \cdot \vec{u}=0$ applies.
If the process starts from rest, the flow must also be
irrotational ($\vec{\nabla} \times \vec{u}=0$)
for all  times and thus we can also write the velocity as
$\vec{u}=\vec{\nabla}\Phi$
where $\Phi$ is the velocity potential which satisfies Laplace's equation
$\vec{\nabla}^2\Phi=0$, following from the continuity equation.
Solving for the flow is just
a matter of solving Laplace's equation, which is linear and
time-independent.
The difficulty in this problem arises via the non-linear dynamic
boundary conditions (the normal stress balance on the interface)
which determines how the interface moves over time.

To describe the implosion of the cylindrical cavity, we naturally use
circular cylindrical coordinates, with the $z$-axis aligned along the
axis of the cylinder, and the surface at $r=R(t)$.
The implosion dynamics is completely $z$-independent, and
two-dimensional because we can consider the implosion of a
circle of radius $R(t)$ in a sheet of water.
If no rotational flow  $u_{\theta}$ is present, the solution
from the Euler equations for the flow field is purely a convergent
radial flow of the form $u_r(r,t)=-Q(t)/r$
and velocity potential $\Phi=-Q(t) \ln (r)$.
This corresponds to the flow of a point-sink with strength $Q(t)$,
which will be determined by the boundary conditions.  The strength $Q(t)$
is related to the areal flux of incoming water by
$\rd A/\rd t= 2 \pi R \dot{R}= 2\pi Q(t)$.
The first boundary condition on
the motion of the interface is the kinematic condition that the interface
moves with a velocity equal to that of the normal velocity of the liquid at
the interface:
\beq
u_{r}(r, t) |_{R(t)}=-\frac{Q(t)}{R}=\dot{R},  \label{kin}
\eeq
showing that the flux $Q(t)=-R\dot{R}$ and the velocity field at any point in
time are completely determined by the radius of the hole and it's velocity,
$u_r(r,t)=R\dot{R}/r$.

To find the evolution of the radius $R(t)$, Euler's equations can
be integrated over the region of the liquid (from $r=R(t)$ to $R_{\infty})$
where $R_{\infty}$ is a far away distance where the radial flow has decayed
or where the inner flow transitions to an outer flow of a different form.
In 2D we must impose this length scale to keep the energy finite because
the potential does not decay as $r\rightarrow\infty$. Experimentally,
this scale will depend on the system size and set-up.
Integrating Euler's equations yields the Bernoulli
integral~\cite{landaufluid} for the pressure in the liquid
$P(r,t)$.  It essentially constitutes a conservation
of energy for each fluid volume. In this case it yields
\beq
P(r,t)=P_{\infty}+\rho \left. \left( \frac{\p \Phi}{\p t}
+\frac{1}{2}|\bg \Phi|^2 \right)\right|_{r}^{R_{\infty}},
\label{pressure}
\eeq
where $P_{\infty}$ is the pressure at $R_{\infty}$ taken as constant.
The normal stress balance at the imploding surface is given by
Laplace's formula for the pressure jump across the interface
$P(R,t)=P_0-\gamma \kappa$, where $P_0$ is the cavity pressure taken
as constant, $\gamma$ is the line tension (analog of surface tension
in 2D), and $\kappa$ is the curvature which is $1/R(t)$ for a circle.
By evaluating the pressure in (\ref{pressure}) at the surface,
we can eliminate it via the normal stress balance and
derive an evolution equation for the radius $R(t)$:
\beq
\Delta P +\frac{\gamma}{R(t)}
=-\rho \left. \left( \frac{\p \Phi}{\p t}
   +\frac{1}{2}|\bg \Phi|^2\right) \right|_{R(t)}^{R_{\infty}}.
\label{bernoulli}
\eeq
In (\ref{bernoulli}) we have incorporated the fact that the
pressure $P_{\infty}$ is higher than $P_0$ by an amount $\Delta P$.
Ignoring the terms in (\ref{bernoulli}) which are $O(R/R_{\infty})$
or smaller and inserting the known form for $\Phi(t)$
yields the well known 2-D analog of the
Rayleigh-Plesset equation
\beq
\Delta P + \frac{\gamma}{R} =\frac{\rho}{2} \dot{R}^2
            - \rho (\dot{R}^2+R\ddot{R})\ln (R_{\infty}/R),
\label{RP}
\eeq
where the dot denotes a time derivative~\cite{plesset77}.

Equation (\ref{RP}) is the second-order non-linear ordinary differential
equation which determines the evolution
of the radius $R(t)$ as it collapses to zero.
Any one dimensional ordinary differential equation of the form
($\ddot{q}+f(q)\dot{q}^2+g(q)=0$),
like (\ref{RP}) above, can be integrated and recast as a
Hamiltonian~\cite{percival}.   We are not surprised by this because
in fact any free boundary problem with inviscid, irrotational,
and incompressible flow is Hamiltonian by nature~\cite{zakharov68}.
However, whereas the Hamiltonian for an arbitrary flow may be quite complicated,
here the Hamiltonian takes a particularly simple form and
represents the total energy per unit length of the system.  The fact that
the dynamical equation (\ref{RP}) can be integrated has been
remarked before~\cite{duclaux07, bergmann06} but the implications of the
integrability had not been fully exploited in this context.
The integrability of the leading order dynamics combined with
the approach to a singularity (where external length and time-scales are
irrelevant) is the origin of the unusual flat mode amplitude
stability spectrum observed in this system~\cite{schmidt08} and other focusing
phenomena like spherical void collapse~\cite{plesset77} or cylindrical~\cite{whitham57} and spherical
shock wave implosion~\cite{evans96}.

The total energy is the sum of the kinetic energy of the water in motion and the
potential energy associated with opening up the hole in the liquid:
\beq
H(R, P_R) = E=\frac{P_R^2}{2 M(R)} +\Delta P\pi R^2 + \gamma\ 2 \pi R, \label{ham}
\eeq
where $R$ is the void radius, $P_R = M(R) \rd R/\rd t$ is the
radial momentum of the implosion, $M(R)= 2 \rho \pi R^2 \ln (R_{\infty}/R)$
is the mass of liquid in motion.  The Hamiltonian is
conserved and thus the total implosion energy $E$ is a constant.
The kinetic term is simply the total
kinetic energy of the sheet of water in motion:
\beq
K.E.=\rho \int_R^{R_{\infty}}u_r^2\: rdr d\theta = \frac{1}{2}M(R) \dot{R}^2
=\frac{P_R^2}{2 M(R)}.  \label {ke}
\eeq
The potential energy due
to the pressure $\Delta P \pi R^2$ comes from the work that must
be done against the external pressure to form the hole and the
surface energy $\gamma 2 \pi R$ is the energy cost for adding
surface length to the hole. One can easily check that this
formulation is equivalent to the 2-D Rayleigh-Plesset
equation (\ref{RP}) by applying Hamilton's equations of
motion ($\dot{R}=\p H/\p P_R$, $\dot{P_R}=-\p H/\p R$).
One technical advantage of using the equation in Hamiltonian,
or integrated form, is that now we no longer need to solve a
2nd order ODE for $R(t)$, but can simply integrate the equation
for $\rd R/ \rd t$, which in fact has known solutions~\cite{duclaux07,bergmann06}.
A more fundamental implication
is that it is now clear that this system
belongs to the class of
integrable dynamics as it has one degree of freedom and
one conserved quantity, the energy.

\section{Asymptotic energy balance and closure dynamics}
For a void which starts from rest, initially the energy is due
completely to the potential contributions from the overpressure
in the liquid and surface tension.  From (\ref{ham}) we
can see that as the void closes, $R\rightarrow 0$, this energy
is transferred into kinetic energy. The asymptotic energy balance is
\beq
E=\frac{1}{2}M(R) \dot{R}^2 =\pi \rho \ln(R_{\infty}/R) (R \dot{R})^2.	
\label{energybalance}
\eeq
A singularity occurs because the closure rate $\dot{R}$
must diverge to compensate for the decreasing area of the void
in order to maintain constant energy. Another way to say this is that
since the mass of liquid in motion $M(R)$ decreases to $0$ at
the moment of disconnection, the closure rate $\dot{R}$ must
diverge.  Both surface tension and the pressure effects
thus become negligible relative to inertia
as the implosion proceeds and $\dot{R}$ diverges.

The leading order dynamics in the inertially dominated regime can be found by
examining the right hand side of (\ref{RP}) or equivalently the energy balance
in Eq. (\ref{energybalance}).  From the energy balance, we see that the
areal flux $Q=-R\dot{R}=\sqrt{E/\pi \rho \ln(R_{\infty}/R)}$ is slowly varying
as $R\rightarrow 0$.
The flux slowly approaches zero as less and less liquid is required to move in
to fill the hole.
Approximating the flux $Q$ as constant, yields the power
law scaling $R=\alpha(t_b-t)^{1/2}$ first proposed by Longuet-Higgins
~\textit{et al} in a study of the sound emission from bubbles released
underwater~\cite{longuet91}.  Here, $t_b$ is the closure time, and
$\alpha$ is related to the flux $Q$ by $\alpha^2=2 Q$ with dimensions of diffusivity
(length$^2/$time). The prefactor $\alpha$ depends on the experimental
specifics. For a bubble released from an underwater nozzle the
appropriate physical scaling is $\alpha^2\sim\sqrt{\gamma R_N /\rho}$, with
$R_N$ the nozzle radius~\cite{keim06}.

In (\ref{RP}) if we include the logarithmic term but treat it
as nearly constant, we find an acceptable power law solution
$R(t)=\alpha (t_b-t)^{\beta}$,
with a weakly varying apparent exponent,
\beq
\beta=\frac{1}{2}\cdot \frac{1}{1-\frac{1}{4\ln (R_{\infty}/R)}}. \label{beta}
\eeq
The exponent $\beta$ tends towards $1/2$
very slowly as $R\rightarrow 0$, but remains slightly greater
than $1/2$, consistent with experimental observations of exponents in
the range $0.54-0.57$~\cite{burton05, bergmann06, keim06, thoroddsen07}.
This is a refinement of the first
estimate of power law of $\beta=1/2$ from~\cite{longuet91}
to include the weak logarithmic dependence on $R$,
consistent with previous analyses~\cite{bergmann06, gordillo05}.

\section{Axial variation}
The above analysis is for the radial implosion of a perfectly
cylindrical void, or equivalently, the implosion of a circular
void in a 2D sheet of inviscid liquid. To extend the results to
make predictions about how the full profile of 3D cavity (like
the nearly cylindrical neck connecting an air bubble to an underwater
nozzle) collapses we make a
long-wavelength approximation for the neck radius so that the
variation in the radius satisfies
$\rd R/\rd z\ll 1$.  This approximation was first used
in~\cite{longuet91} with the leading
order dynamics treated as quasi-2D, and follows the above analysis
for a cylinder but with $z$ simply treated as a parameter.
The flow at each height converges radially with flux $Q$,
and is isolated from any coupling in the $z$-direction (there is no
energy or momentum transfer along $z$).  For simplicity, we
first examine the result ignoring logarithmic corrections so that
the flux is constant in time (corresponding to the
$R=\alpha \tau^{1/2}$ power law) and in $z$.

We initialize the surface at some $t=0$ within the inertially dominated
implosion regime.  In the region near the neck minimum the shape is quadratic
\beq
R(z,t=0)=H(z)=H_{0}+ c_0 z^{2}.  \label{ic}
\eeq
To stay within the quasi-2D limit the aspect ratio is
constrained to be small, $H_0/L =\sqrt{c_0 H_0} \ll 1$,
where we have used the axial length scale $L=\sqrt{H_0/c_0}$.
Taking the approximation that the flux is constant,
we can write the evolution of the profile as
\beq
R(z,t)=\alpha (t_{b}(z)-t)^{1/2},
\eeq
with the only difference between the evolution of the radius at
difference heights being the implosion time, $t_b(z)$, which
is set by the initial radius profile.
The implosion time is the time when the radius at that height
will collapse to zero as determined by the initial size (\ref{ic}):
\beq
t_{b}(z)=\left( \frac{H(z)}{\alpha} \right)^{2}.
\eeq
With nothing to break the up/down symmetry about $z=0$, the
height which starts with the minimum radius $H_0$,
will be the height of the disconnection event at $t=t_{\ast}=t_b(0)$.
Writing the shape evolution in terms of the time until
disconnection $\t=t_{\ast}-t$ yields
\beq
R(z,\t)=\alpha \left(\left(\frac{H(z)}{\alpha}\right)^{2}
  -\left(\frac{H(0)}{\alpha}\right)^{2} +\t  \right)^{2}.
\eeq
By proper rescaling of the axial and radial coordinates,
the evolution has a self-similar form.  Taking $\zeta=R/h_{min}$ and
$\eta=z/h_{min}$ with $h_{min}=R(z=0,t)$, the shape in
the rescaled coordinates is approximated by:
\beq
\zeta=(2 c_0 H_0 \eta^{2}+1)^{1/2}.  \label{similar}
\eeq
In the limit $h_{min}\rightarrow 0$,
the shapes becomes conical ($R(z)=\sqrt{2 c_0 H_0}\ z$)
as seen in the experiments (Fig.~\ref{bubbles}) and shown in~\cite{longuet91}.
The relevant axial length scale is proportional to $h_{min}$, so that
it decreases in pace with the radial scale and the
aspect ratio of the shapes are maintained.
Of importance is the
dependence of the self-similar form on the initial aspect ratio $\sqrt{c_0 H_0}$.
The final slope of the cones is directly dependent on the initial aspect
ratio of the quadratic neck shape. An initially slender shape will
remain slender and the long-wavelength approximation analysis is self-consistent
in this respect.  

Accounting for the logarithmic corrections to the power law
exponent $\beta$ in Eq.~\ref{beta}, we find a self-similar form
\beq
\zeta=(b\eta^{2}+1)^{\beta},  \label{similar2}
\eeq
with $\eta=z/h_{min}^{1/2\beta}$ and $b=c_0 H_{0}^{1/\beta-1}/\beta$.
The region where the shape is approximated by this form
is bounded by
\beq
\eta\ll \sqrt{\frac{H_{0}}{c_0(\frac{1}{\beta}-1)h_{min}^{1/\beta}}},
\eeq
which grows as $h_{min}$ shrinks to $0$. This form assumes that nearby
layers evolve with the same $\alpha$ and $\beta$, a good approximation
 for slowly varying shapes because $\alpha$ and $\beta$ themselves
have only logarithmic dependencies on the local radius $R(z,t)$.  Accounting
for the weak $z$-variation of $\alpha$ and $\beta$,
and solving the governing equations
self-consistently replacing $R_{\infty}$ with the axial length scale yields
slight corrections to the exponent~\cite{egg07}.

Now, with the logarithmic corrections included, the axial
length scale is proportional to $h_{min}^{1/2\beta}$
which decreases almost in pace, but a bit slower, than the radial scale.
For a power law measured over a finite range of time we can write
$h_{min}\propto \tau^{1/2+\delta}$ so that the axial scale $\propto \tau^{1/2-\delta}$,
where $\delta$ measures the difference in the exponent from $1/2$,
$\delta=1/4\ln(R_{\infty}/h_{min})$.
In other words, if the radius is decreasing with a power slightly greater
than $1/2$ (as observed in the experiments), then the quasi-2D model predicts
the axial length scale will decrease with a power
equally less than $1/2$. This result is qualitatively consistent
with experimental measurements of the axial radius of curvature at the
minimum~\cite{bergmann06, thoroddsen07} which find powers less than but
still close to $1/2$.

It is worthwhile to note that although the profiles have a self-similar
form, it does not mean that the dynamics here is universal.  Though
often times universal dynamics exhibit self-similarity, in this case
the reverse is not true.  The underlying dynamics here is in fact nearly
integrable, with the implosion energy at every height $E(z)$ conserved.

When extending the 2D-cylindrically symmetric model to capture
the evolution of axisymmetric 3D shapes,
we must also modify the Laplace pressure contribution due to
surface tension at the interface which appears in the normal stress
balance.  Now, the surface is no longer a curve
 in a plane, but has an axial curvature component. Because the
surface is curved outwards in the axial direction, it works in opposition
to the azimuthal inwards curvature, effectively reducing the Laplace
pressure jump across the interface.  Directly from above, we know that
the axial curvature at the minimum will increase at roughly the
same rate as the radial curvature.  This fact will later allow us to simply
include the effect of the axial curvature by reducing the Laplace pressure
by a fractional amount when comparing to the experiments over a limited range.
We can write the axial radius of curvature $R_{ax}=\chi h_{min}$,
where $\chi$ is a constant determined by the initial
aspect ratio from (\ref{similar}).  Thus, a more realistic
value of the Laplace pressure jump
\beq
\gamma \kappa=\gamma (1/h_{min}-1/R_{ax})=\gamma(1-1/\chi)/h_{min} \label{laplace}
\eeq
can be used in (\ref{bernoulli}).

The quasi-2D model outlined above provides a simple way to interpret and analyze
the experimental results available today. It contains all the ingredients
necessary to explain experimental observations of the time dependence
of cavity profiles during implosion.  We will see it is more than sufficient for the
purposes of the present paper,
not to predict the complete details of the asymptotic approach to the singularity,
but to examine the effect of azimuthal asymmetry on the implosion dynamics
within the experimental regime.

\chapter{Azimuthal perturbations}
\label{aziperts}
With the axisymmetric implosion dynamics in the limit of an infinite cylinder
now determined by equation (\ref{RP}) or conservation of the
Hamiltonian (\ref{ham}), we can begin to investigate the stability of the implosion
under a slight azimuthal shape distortion which destroys the symmetry.
Because the underlying leading order dynamics is integrable and possesses
a natural memory mechanism via conservation of the energy, we will
see the stability spectrum takes a special form.  No single wavelength perturbation
will grow fastest to overtake the others, as typically occurs in
interfacial instabilities.  Instead, all the modes will grow at the same rate,
logarithmically slow relative to the rapid collapse of the mean radius.
This unique property is the result of the combination of the approach
towards a singularity and the integrable dynamics~\cite{schmidt08}.
Since all modes grow identically, the initial spectrum of mode amplitudes
for a generic distortion is frozen in time as the singularity is approached,
encoding an infinite amount of information about the initial state
in form of vibrational modes about the radial implosion.

Representing a distortion to the shape as the sum over the Fourier
modes $\cos(n \theta)$ shifts the surface to lie at
\beq
r=S(\theta,t)=R(t)+\sum_n b_n(t)\ \cos(n\theta),
\eeq
where the $b_n(t)$ are the mode amplitudes and assumed to be
small relative to mean radius $R(t)$ ($b_n(t)/R(t)\ll 1$).
The velocity potential and velocity field in the liquid
depend on the boundary conditions on the surface, so distorting the
surface causes a disturbance flow in the surrounding liquid.
The leading order and disturbance flow have a total velocity potential of the form:
\beq
\Phi (r, \theta, t)=Q(t)\ln (r)+\sum_n d_n(t) r^{-n}\cos (n\theta). \label{potential}
\eeq
The kinematic condition at the surface (\ref{kin}) at $O(a_n/R)$
determines the coefficients
\beq
d_n(t)=-\frac{R^{n+1}}{n}\left(\dot{b}_n+b_n\frac{\dot{R}}{R}\right). \label{cn}
\eeq
Note that here, the azimuthal components $u_{\theta}$ and $n_{\theta}$ caused
by the disturbance
did not enter into the kinematic equation (\ref{kin}) because they are both
of $O(b_n)$, thus making their product $O(b_n^2)$.
Applying the Bernoulli integral (\ref{bernoulli}), and
the Laplace formula for the pressure jump across the surface,
also to $O(b_n)$ yields the evolution equation for the mode
amplitudes with coefficients which depend on the radial implosion dynamics:
\beq
\ddot{b}_n+\left(\frac{2\dot{R}}{R}\right)\dot{b}_n
          +\left(\frac{\ddot{R}}{R}(1-n)\right)b_n=0.
\label{agov}
\eeq
We can take advantage of the fact that we have a conserved energy
to use (\ref{RP}) and (\ref{ham}) to rewrite
$\dot{R}$ and $\ddot{R}$ in terms of $E$ and $R$.  This,
along with using $\rd/\rd t=\dot{R}\:\rd/\rd R$ allows us to
rewrite (\ref{agov}) so that the independent variable is $R$
rather than $t$.
Taking the limit that the implosion is
inertially dominated and the radius has decreased far away from
the system size $R_{\infty}$, yields
\beq
b_n''+\!\left(\!1+\frac{1}{2\ln(\frac{R_{\infty}}{R})}\right)\!\frac{b_n'}{R}
+(n-1)\!\left(\!1-\frac{1}{2\ln(\frac{R_{\infty}}{R})}\right)\!\frac{b_n}{R^2}
= 0,
\label{agrowth}
\eeq
where $'$ denotes a derivative with respect to $R$.

Immediately we can clearly see the amplitude freezing and vibrational
behavior by taking the limit
$\ln(R_{\infty}/R)\gg $ in (\ref{agrowth}), which corresponds to a
radial implosion follow $R=\alpha \tau^{1/2}$.  Then,
the governing equation takes the particularly simple equidimensional form
\beq
b_n''+\frac{b_n'}{R}+(n-1)\frac{b_n}{R^2} = 0,
\eeq
with solutions for $b_n(R)$ of the form $b_n\propto R^{\pm \rm i \sqrt{n-1}}$.
These solutions neither grow nor decay because
the eigenvalues are entirely imaginary,
but instead oscillate with constant amplitude $b_{n,0}$ and an $n$-dependent
frequency.  The real part of the vibration can be written
\beq
b_n(R)=b_{n,0}\cos(\sqrt{n-1}\ln R + \phi_{n,0}),  \label{vibrate}
\eeq
with $\phi_{n,0}$ an arbitrary phase shift defined by the
initial conditions.  The frequency of the vibrations, $\sqrt{n-1}/R$ is
$n$-dependent, with higher modes vibrating faster.
All the vibrations become faster
and faster, or ``chirp'', as $R\rightarrow 0$ due to the divergent
$1/R$ dependence.  This is a direct consequence of the singularity
formation, where no other length or time-scales other than the closure
radius $R$ and time $\tau$
are involved in the dynamics.  Thus, from simple scaling the frequency
is forced to be proportional to $1/R$.

\begin{figure}
\psfrag{aaaa}{\large{$b_n/b_0$}}
\psfrag{RRRR}{\large{$R/R_0$}}
\begin{center}
\includegraphics[width=0.8\textwidth, angle=0]{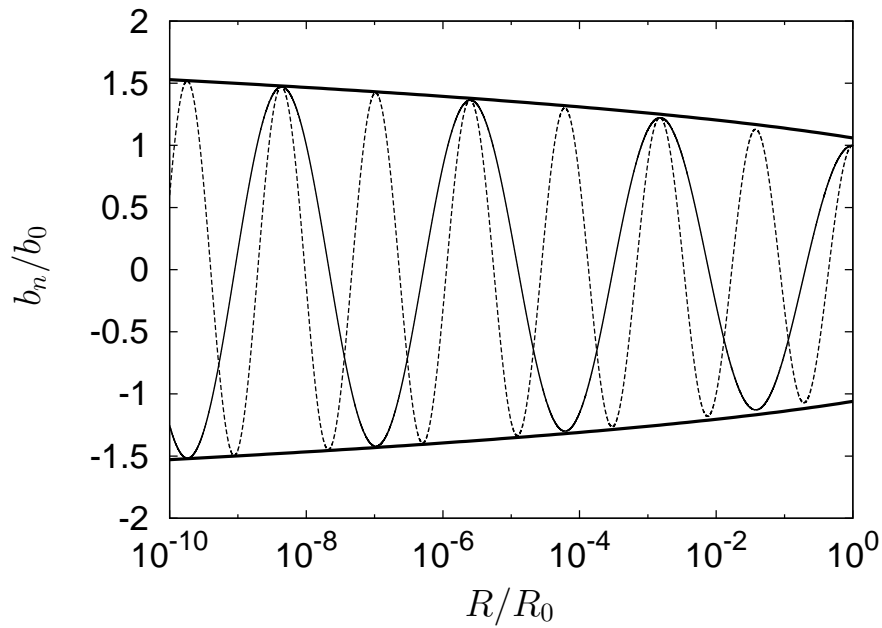}
\end{center}
\caption{ Numerical solutions to (\ref{agov}) for vibrational mode amplitudes
 in inertial limit, for modes $n=2$ (dashed line), $n=5$ (dotted line).
 The amplitudes grow weakly as the mean radius decreases, like
 $(\ln(R_{\infty}/R))^{1/4}$, predicted from solution to (\ref{agrowth})
 using (thick solid line). Mode amplitudes are rescaled by their initial size
 $b_0$ and the mean radius $R$ is rescaled by its initial size $R_0$, and
 $R_{\infty}/R_0=100$.}
\label{inertialmodes}
\end{figure}

To check that including the logarithmic corrections does not qualitatively
change the vibration freezing and chirping characteristics, we analyze (\ref{agrowth})
by introducing a second
slowly varying length scale defined as $y=\ln(R_{\infty}/R)$.
Performing a standard multiple scale analysis~\cite{bender}
we find that the modes are primarily oscillatory with amplitudes that
grow extremely weakly and uniformly, as $(\ln(R_{\infty}/R))^{1/4}$.
This growth is shown in Fig.~\ref{inertialmodes} and envelopes the numerical
solutions to (\ref{agov}) for $n=2$ and $5$.

The growth can also be shown by
transforming (\ref{agrowth}) into the $y$ variable:
\beq
\frac{\rd^2 b_n}{\rd y^2}-\frac{1}{2 y}\frac{\rd b_n}{\rd y}
     +(n-1)\left(1-\frac{1}{2y}\right)b_n=0. \label{yeqn}
\eeq
Solutions to (\ref{yeqn}) are of the form
$b_n(y)=y^{3/2}e^{-i\sqrt{n-1}\ y}f(y)$  with $f(y)$ satisfying
\beq
2 y \frac{\rd^2 f}{\rd y^2}+\frac{\rd f}{\rd y}(5-4i\sqrt{n-1} y)
  +f(1-n-5i\sqrt{n-1})=0.
\eeq
Again transforming to the complex variable $z=2i\sqrt{n-1}\ y$
we arrive at
\beq
z\frac{\rd^2 f}{\rd z^2}+\left(\frac{5}{2}-z\right)\frac{\rd f}{\rd z}
  -\left(\frac{5-i\sqrt{n-1}}{4}\right)f=0
\eeq
which is of the form of the confluent hyper-geometric differential
(Kummer's) equation.  The solution for $f(y)$ when
$\rm{Re}(z)\!\gg\! 1$ (or~$y\!\gg\!1$,~$R_{\infty}/R\!\gg\!1$)
decay like $f \propto z^{-5/4}$.
Taken together with the above form for $b_n(y)$ yields
a weak growth in the amplitude like
$b_n(y)\propto y^{3/2 - 5/4}=y^{1/4}$ or $b_n(R)\propto (\ln(R_{\infty}/R))^{1/4}$.

The amplitude growth is so weak (in Fig.~\ref{inertialmodes},
increasing only by a factor of $1.5$ as the
mean radius deceases by 10 orders of magnitude) that the amplitudes are
effectively constant. More importantly, because all modes grow
within the same $n$-independent envelope, an initial mode distribution
will still be preserved as the implosion proceeds.
Numerical solutions to (\ref{agrowth}) for $n=2$ and $5$ are shown
in Fig. ~\ref{inertialmodes}.  The weak growth follows the
analytic prediction of $\ln(R_{\infty}/R)^{1/4}$.

\begin{figure}
\begin{center}
\includegraphics[width=0.6\textwidth, angle=0]{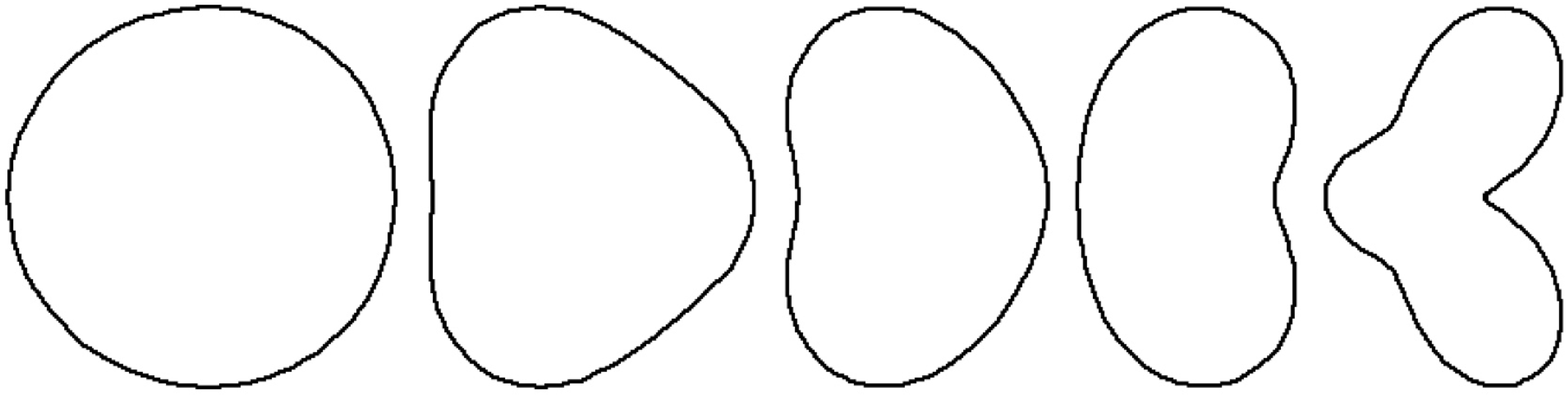}
\end{center}
\caption{Illustration of void shapes for superposition of
  vibrational modes $n=2$ and $3$, with dynamics given by Eq.~\ref{vibrate} in text.
  The shapes are rescaled by the mean radius
which is collapsing to zero (left to right).  As the mean radius collapses,
the shapes become more distorted.  Because the $n=2$ and 3 modes are not
vibrating in phase, complex shapes ensue.}
\label{shapes}
\end{figure}

Considering the vibrational modes relative to the leading-order
radial implosion the shape becomes more distorted as $R$
decreases to $0$.  Though the amplitudes are constant, the
relative amplitudes $b_n/R$ grow like $\tau^{-1/2}$,
causing the shape to become increasingly more distorted.
Taking the linear stability results to the
point when $b_n=O(R)$ predicts the development of cusp-like structures
on the interface.  The shapes, especially when two or more
modes are present can become quite complex as the vibrations occur
with $n$-dependent frequencies.  An illustration of shapes that occur as the
void collapses for a combination of modes $n=2$ and 3
is shown in Fig.~\ref{shapes}.

At some point (once the mean radius $R$ has collapsed to the size
of the largest mode amplitude present)
the approximation $b_n/R \ll 1$ breaks down and non-linear
effects cannot be ignored.
Recent numerical results utilizing a conformal
mapping method have explored the non-linear regime that follows and found
a rich landscape of different types of pinch-off depending sensitively
on the initial conditions~\cite{kostyapreprint}.

These vibrational modes are purely inertial, unlike
most vibrations or waves on fluid interfaces.  For example,
a jet of water issued from an elliptic shaped nozzle will also
undergo vibrations of the cross section shape, but this
effect is due to a competition between inertia and
surface tension~\cite{rayleigh79b}.  The inertial shape vibrations here
occur because once the flow and pressure fields are perturbed, there is no
mechanism for the extra energy associated to be dissipated in this closed and
conservative system.  The extra energy could go into or out of
the main radial implosion, but in this situation it
remains in the perturbation leaving the radial dynamics unchanged.


To help visualize the dynamics of the inertial vibrations, we can
track how the fluid moves in and out of the wells around the circle.
The corresponding corrections to the pressure field and velocity
field (after subtracting out the leading order radial implosion) can be
calculated using the potential from (\ref{potential}) with the
appropriate coefficients from (\ref{cn}) given the evolution of
the mode amplitude from (\ref{agov}).  The disturbance
pressure field  $P(r,\theta)$ is found from evaluating the Bernoulli
integral (\ref{bernoulli}) at $O(b_n)$ and yields:
\beq
\delta P(r,\theta,t)=
\frac{\rho \cos (n\theta)}{n}\!\left(\!\frac{R}{r}\right)^n
\!\left(\!R\dot{s_n}\!
+\! s_n\dot{R}\left(1\!+\!n\!-\!nR^2/r^2\right) \right),
\label{dP}
\eeq
where $s_n=\dot{b}_n+b_n\dot{R}/R$.
The disturbance velocity field is $\vec{\nabla} \Phi$ at $O(b_n)$:
\beq
\delta\mathbf{u}(r,\theta,t)
=\left(\dot{b}_n+b_n\frac{\dot{R}}{R}\right)\left(\frac{R}{r}\right)^{n+1}
(\cos (n\theta)\, \hat{r}+\sin (n\theta)\, \hat{\theta}).
\label{du}
\eeq

\begin{figure}
\begin{center}
\includegraphics[width=0.5\textwidth, angle=0]{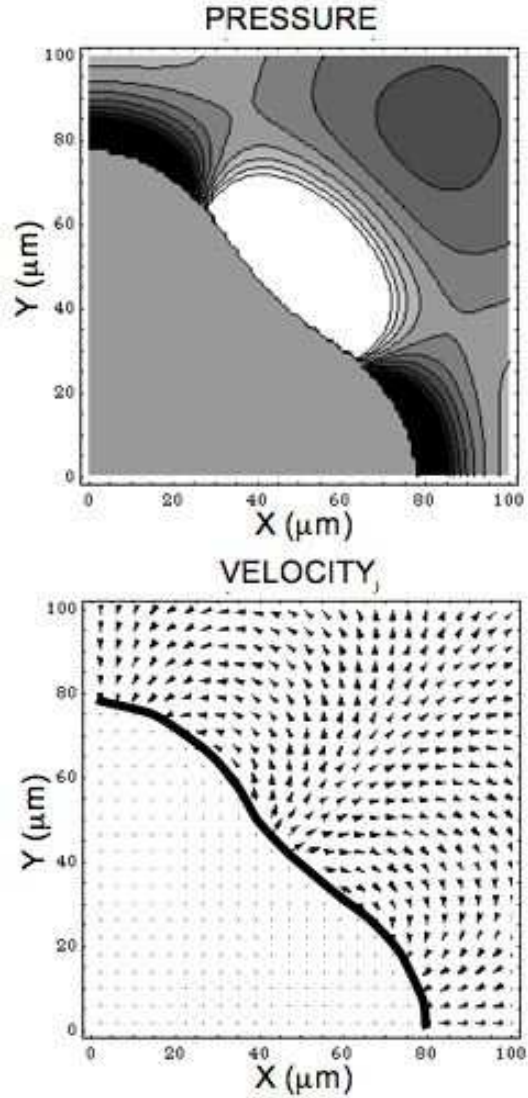}
\caption{Example of disturbance to the pressure and velocity fields outside the
surface due to an $n=4$ shape distortion (one quadrant shown). For the pressure
contours, inside the void the disturbance
pressure is zero (medium grey) and outside the pressure has regions of positive (lighter)
and negative (darker) pressure. The surface location is highlighted by the thick
black line in the lower plot for the velocity.
At this instant in time, the mean radius $R=70\mu$m
  and has an implosion speed $\dot{R}=-0.7\mu$m$/\mu$s.
  For clarity, a large perturbation amplitude is shown with
  $b_4=8\mu$m and $\dot{b_4}=-0.1 \mu$m$/\mu$s.}
\end{center}
\label{fields}
\end{figure}

For an $n=4$ distortion the disturbance pressure and velocity fields are
shown in Fig.~\ref{fields}.
Physically, the pressure builds up in
the wells, leading to a disturbance flow outwards towards the peaks.
The pressure and velocity field oscillate, but are out of phase,
so that as the shape rounds up to a circle, the pressure is uniform
around the circle (in $\theta$) but flow continues, and creates a
well. The pressure in the well begins to build
up again (lightest region in Fig.~\ref{fields}),
and causes a reversal of flow out of the wells.

Thus far we have shown that when a small shape distortion is
introduced,  an imploding cylindrical void will undergo
inertial shape vibrations about the radial implosion dynamics.  Within the
linear regime, these vibrations have constant amplitudes but grow relative
to the shrinking mean radius.  This scenario naturally provides an
explanation for why experiments observe a wide variety of different
disconnection events --- the final moments at disconnection are sensitively
linked to the earliest moments in the implosion process.
Below we show that including
additional physical effects into the model do not destroy this unique
behavior within the experimental regime.

\chapter{Consequences of additional physical effects}
\label{phys}
The inertial shape vibrations were analyzed in a specialized
limit of flow and fluid properties.  It was assumed that the leading
order dynamics was a purely radial implosion, that the liquid was completely
inviscid, and was analyzed in the far asymptotic limit when the dynamics was
completely inertially dominated.  For the purely radial implosion in the inviscid
liquid, the dynamics was shown to be integrable and evolve towards a singularity.
Below we will test the consequences on the vibrational modes
of destroying these two properties, in ways that are physically relevant.

This
is done by first including viscous dissipation in the external liquid in the
simplest manner possible, which destroys the integrability because the collapse
energy is no longer conserved.  Second, we remove the singularity by introducing a
weak rotational flow into the external liquid which prevents the hole from
collapsing to a point and forming a finite time singularity.  Both of these
effects are physically relevant as well -- no experiment uses a perfectly inviscid
liquid or can completely eliminate the presence of a weak swirling flow.
We will find that within the experimental regime, the inertial shape vibration
behavior described in the preceding section will be robust against these effects,
for the small mode numbers (large wavelength disturbances) explored in the experiments.
Finally, even though surface tension becomes irrelevant asymptotically, we account
for capillary effects and find the vibration frequency is increased in the experimental
regime before the vibrations settle in to their asymptotic behavior.

\section{Viscous dissipation}
To include viscous dissipation we must first consider the effect
on the leading-order radial implosion dynamics, and then consequently
the effect on the vibrational mode behavior.
When viscous dissipation is included in the model, the implosion
energy $E$ is no longer strictly conserved and the leading-order symmetric
collapse dynamics are no longer integrable and Hamiltonian. However,
Eq.~(\ref{RP}) still holds once the
viscous contribution to the normal stress balance is taken into account.
For the leading-order radial flow, viscous dissipation does not
alter the form of the equations governing fluid flow but it
does contribute to the normal stress balance
on the surface:
$p(R)=p_0-\gamma \kappa +2 \mu \p u_r /\p r$, where $\mu$ is the
liquid viscosity.
Since $\p u_r/\p r=-\dot{R}/R$ on the surface, the governing equation
(\ref{RP}) is then modified to
\beq
\Delta p + \frac{\gamma}{R} + 2 \mu\frac{\dot{R}}{R}=\frac{\rho}{2} \dot{R}^2
            - \rho (\dot{R}^2+R\ddot{R})\ln (R_{\infty}/R). \label{RPvisc}
\eeq
Comparing the inertial and viscous terms we see that viscous dissipation
will be relevant to the process
when $\dot{R}{R}/\nu = Q/\nu= O(1)$ (with the kinematic viscosity $\nu=\mu/\rho$),
i.e., when the dynamic Reynolds number is $O(1)$.
If the implosion begins in an inertially dominated regime where $E\approx$constant,
this occurs very late in the implosion,
at a length scale beyond the reach of a continuum theory, seen from
$Q=\sqrt{E/\pi \rho \ln(R_{\infty}/R)}$.  For the air cavity in water
experiments the dynamic Reynolds numbers are typically on the order
of $10^{2}-10^3$, which means that within the experimental regime, the radial
collapse is inertially dominated and $E=$constant will produce the
observed radial implosion dynamics.

Once we consider the shape perturbations, the velocity field is no
longer purely radial, but has an azimuthal ($\theta$) component.  To
fully include the effect of a small viscosity in this problem would
require a complete change in the solution method.  The presence of viscosity
means that vorticity can be generated and therefore the flow is no
longer truly potential.  However, if the vorticity is confined
to a very shallow boundary layer near the surface where it is generated,
we can again approximate the viscous effects as only contributing to the normal
stress balance.  Then, the stability analysis yields a revised equation for
$b_n(t)$:
\begin{eqnarray}
\ddot{b}_n &+&\left(\frac{2\dot{R}}{R}+ 2\nu \frac{n(n+1)}{R^2}\right)\dot{b}_n
  \nonumber \\
  &+&\left(\frac{\ddot{R}}{R}(1-n)+2\nu\frac{n(n-1)\dot{R}}{R^3}\right)b_n=0.
  \label{avisc}
\end{eqnarray}

\begin{figure}[]
\subfigure[][\ $n=2$]{
  \psfrag{aaa2}{$b_2/b_0$}
  \psfrag{RRRR}{$R/R_0$}
\includegraphics[width=0.5\textwidth, angle=0]{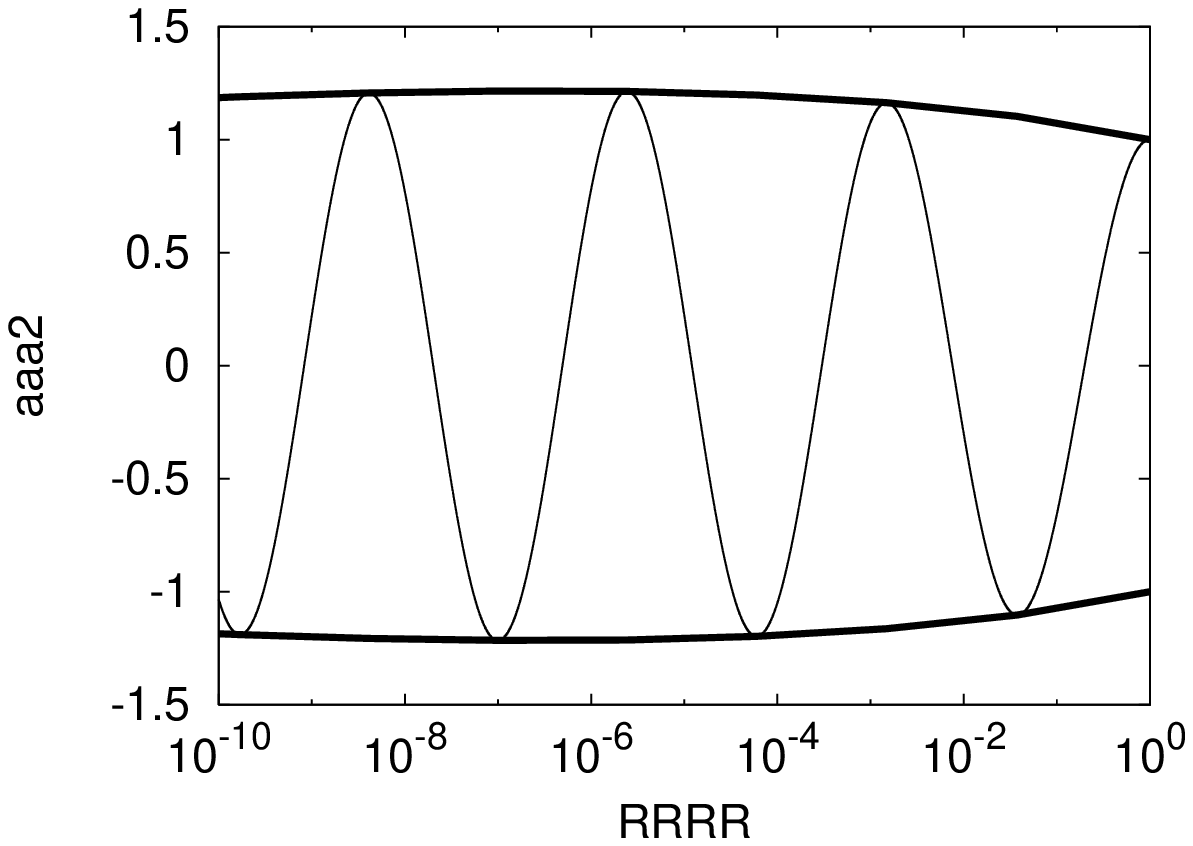} }
\subfigure[][\ $n=2-10$]{
  \psfrag{aaaa}{$b_n/b_0$}
  \psfrag{RRRR}{$R/R_0$}
\includegraphics[width=0.5\textwidth, angle=0]{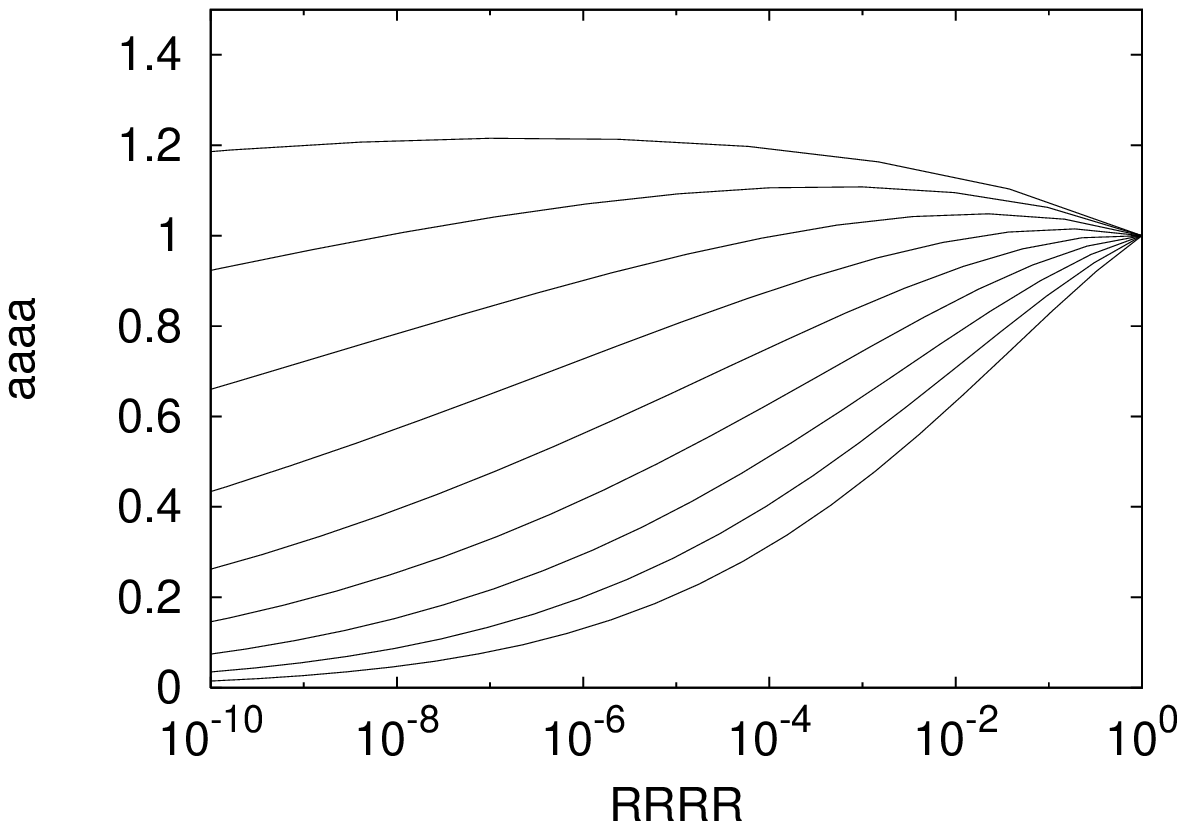} }
\subfigure[][\ spectrum]{
  \psfrag{aaaa}{$b_n/b_0$}
  \psfrag{n}{$n$}
  \psfrag{xxxx2}[tc][cl]{\footnotesize{$R/R_0=10^{-2}$}}
  \psfrag{xxxx5}[tc][bl]{\footnotesize{$R/R_0=10^{-5}$}}
\includegraphics[width=0.5\textwidth, angle=0]{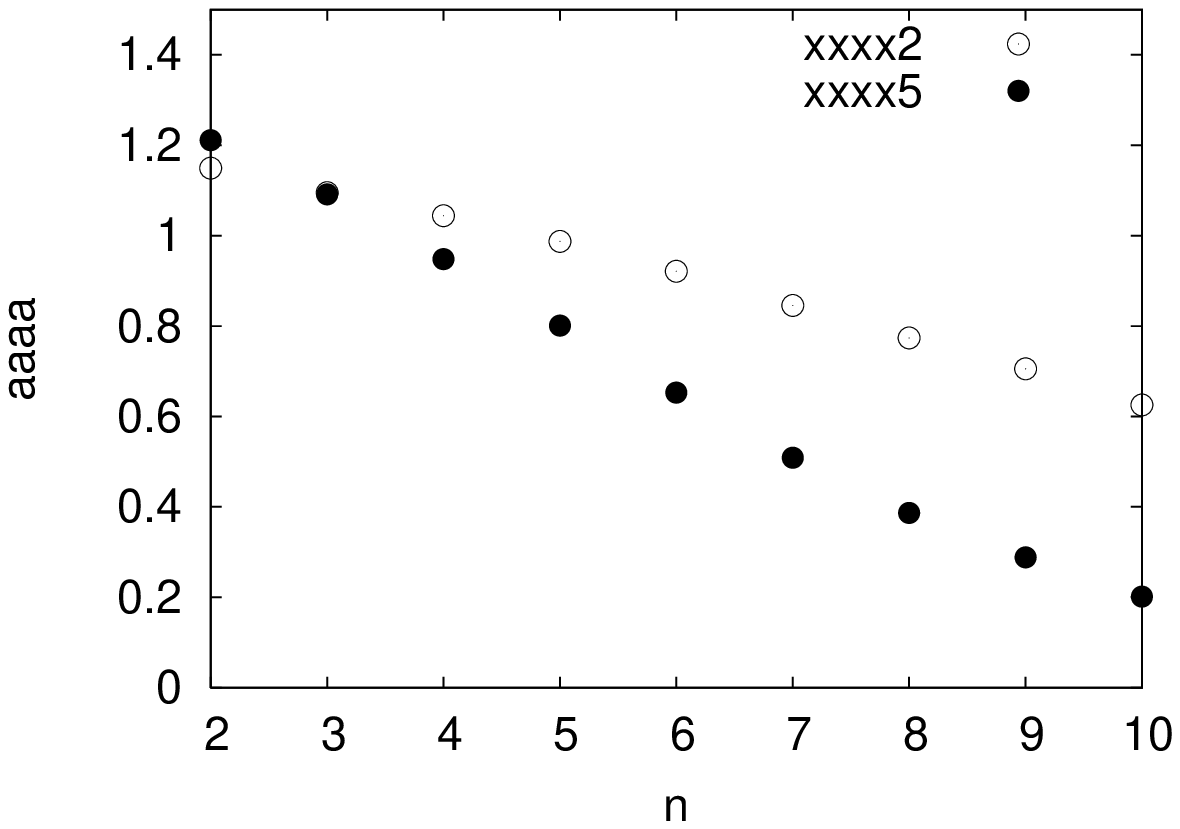} }
\caption{Amplitude decay due to viscous dissipation at the interface.
  Here, $R_{\infty}/R_0=10^{2}$ and the dynamic Reynolds number for the flow
  is $Q/\nu=10^3$ when $R=R_0$.
The amplitudes are rescaled by their initial values $b_0$ and the mean radius
is rescaled by its initial value $R_0$.
(a) Representative vibration for $n=2$ (thin line) shows that viscous dissipation
 causes a decrease in the amplitude (envelope is thick line)
  and frequency of vibration.
(b) Envelope of vibrational mode amplitudes for $n=2-10$ (upper- to lower-most).
    Short wavelength (high $n$) disturbances are efficiently smoothed out
    by the presence of viscosity.
(c) Amplitude spectrum taken at $R/R_0=10^{-2}$ (open circles)
    and at $R/R_0=10^{-5}$ (solid circles).}
\label{visc_2}
\end{figure}

\begin{figure}[]
\subfigure[][\ $n=2$]{
  \psfrag{aaa2}{$b_2/b_0$}
  \psfrag{RRRR}{$R/R_0$}
\includegraphics[width=0.5\textwidth, angle=0]{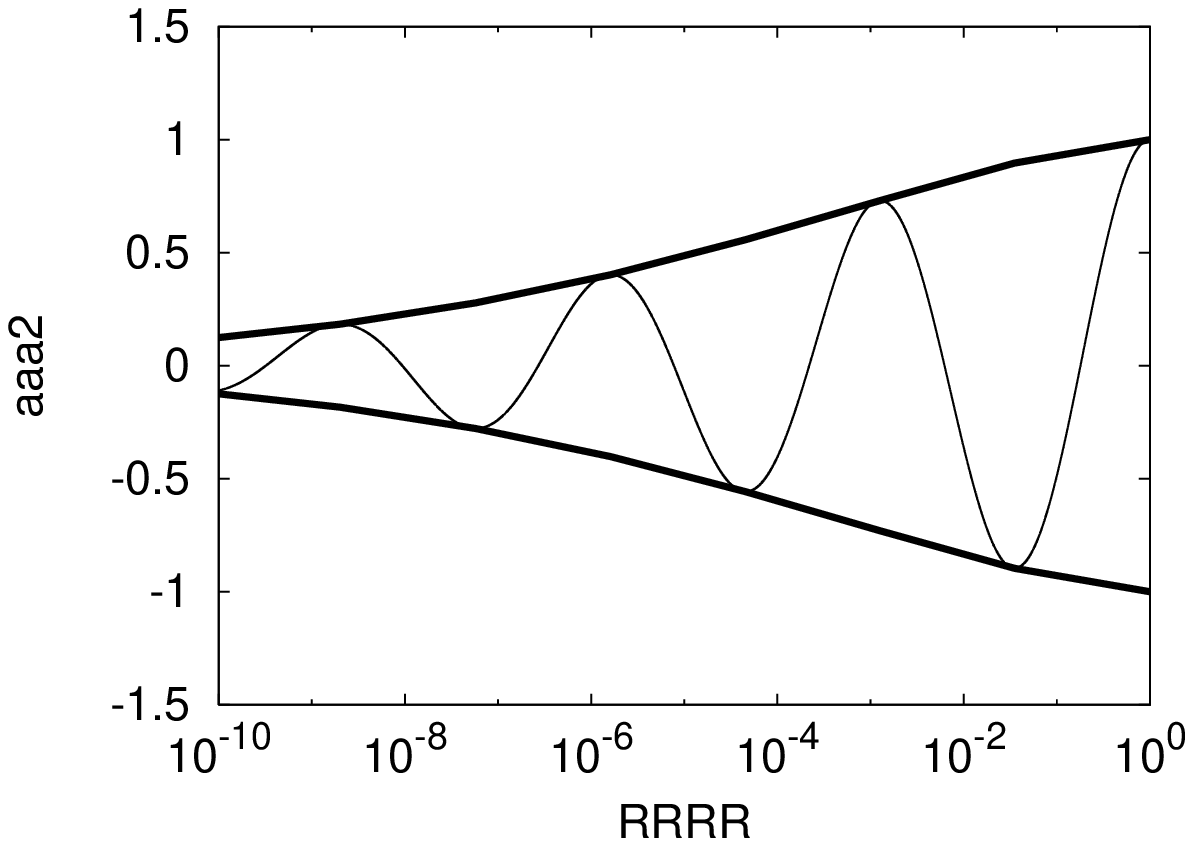} }
\subfigure[][\ $n=2-10$]{
  \psfrag{aaaa}{$b_n/b_0$}
  \psfrag{RRRR}{$R/R_0$}
\includegraphics[width=0.5\textwidth, angle=0]{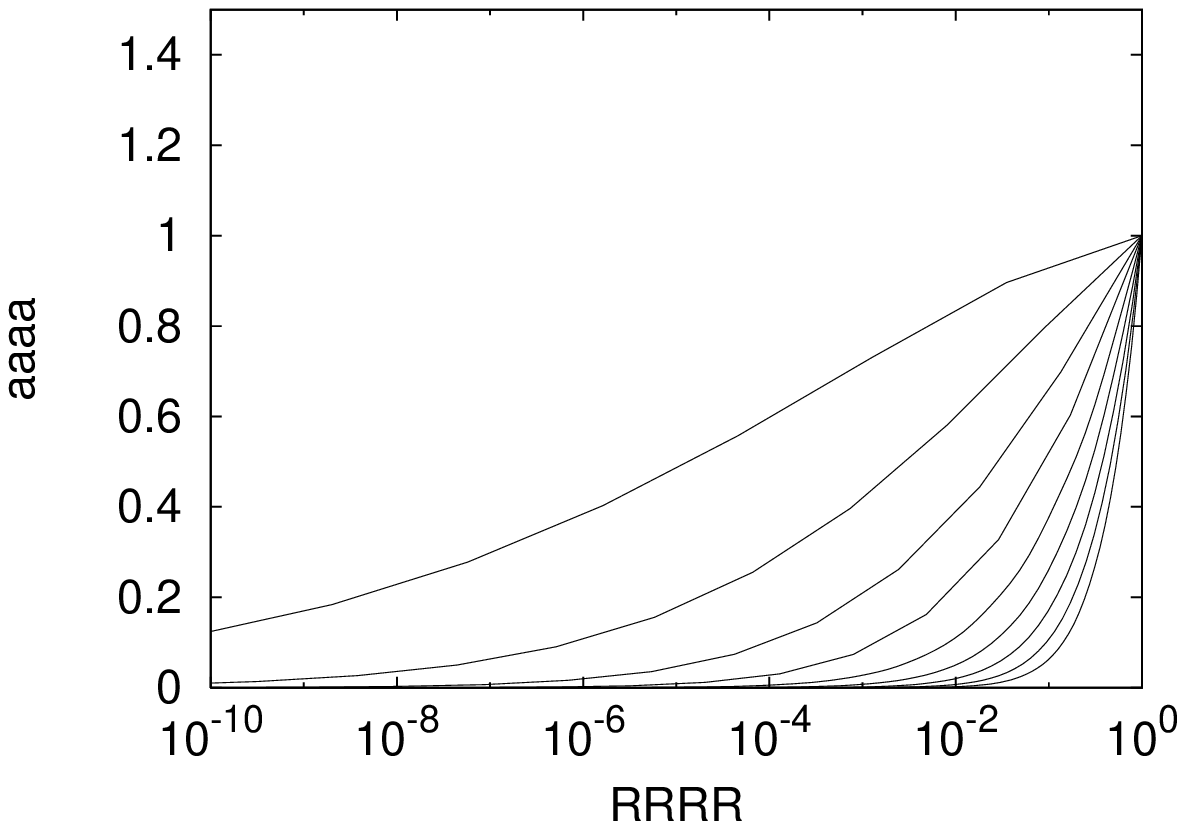} }
\subfigure[][\ spectrum]{
  \psfrag{aaaa}{$b_n/b_0$}
  \psfrag{n}{$n$}
  \psfrag{xxxx2}[tc][cl]{\footnotesize{$R/R_0=10^{-2}$}}
  \psfrag{xxxx5}[tc][bl]{\footnotesize{$R/R_0=10^{-5}$}}
\includegraphics[width=0.5\textwidth, angle=0]{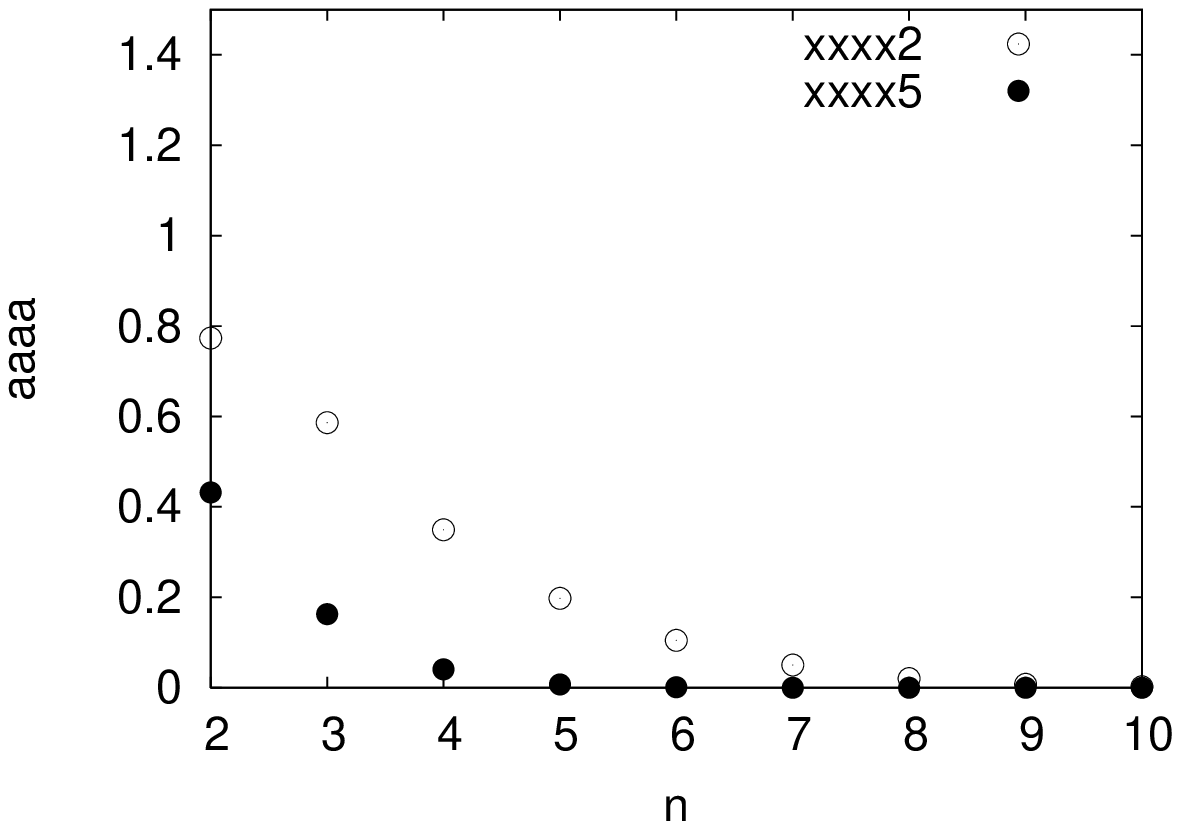} }
\caption{Amplitude decay due to viscous dissipation at the interface.
  Here, $R_{\infty}/R_0=10^{2}$ and the dynamic Reynolds number for the flow
  is $Q/\nu=10^2$ when $R=R_0$.
The amplitudes are rescaled by their initial values $b_0$ and the mean radius
is rescaled by its initial value $R_0$.
(a) Representative vibration for $n=2$ (thin line) shows that viscous dissipation
 causes a decrease in the amplitude (envelope is thick line)
  and frequency of vibration.
(b) Envelope of vibrational mode amplitudes for $n=2-10$ (upper- to lower-most).
    Short wavelength (high $n$) disturbances are efficiently smoothed out
    by the presence of viscosity.
(c) Amplitude spectrum taken at $R/R_0=10^{-2}$ (open circles)
    and at $R/R_0=10^{-5}$ (solid circles).}
\label{visc_1}
\end{figure}

From (\ref{avisc}) we can see that viscous effects can now play an earlier role for high
mode numbers, because the viscous terms have prefactors $O(n^2)$ and will
quickly begin to dominate.
Figs.~\ref{visc_2} and~\ref{visc_1} plot results of numerical solutions
to (\ref{avisc}) for two values of dynamic Reynolds number, $10^2$ and $10^3$ (which over the range
of the data change by only a factor of about $2$). Figs.~\ref{visc_2}
and~\ref{visc_1} (A) plot the vibration for a single mode $n=2$, showing a decay in
the amplitude relative to inviscid case (see Fig.~\ref{inertialmodes}), but the
chirping behavior of the frequency is unchanged.  In Figs.~\ref{visc_2} and~\ref{visc_1}
(B), the envelopes for modes $n=2-10$ are plotted, and clearly show the $n$-dependence
of the rate of decay. All mode amplitudes experience decay due to viscous dissipation,
but the short wavelength (high $n$) feel the strongest effect.
The reason for the apparent amplitude growth in Fig.~\ref{visc_2} B for $Q/\nu=10^3$
is simply that the inertially dominated vibrations have
weak growth (as in Fig.~\ref{inertialmodes}) which acts until viscous dissipation
takes over as $R\rightarrow 0$. For the smaller $Q/\nu=10^2$ in Fig.~\ref{visc_1},
the weak growth is completely washed out by a rapid decay in the amplitude.

To quantify the decay, we plot the spectrum of the mode amplitudes as a function
of mode number $n$ at a given mean radius.  We choose two values of the mean radius at
which to evaluate the spectrum, $R/R_0=10^{-2}$ and $10^{-5}$ (so the mean radius collapses
2 or 5 decades from its initial value). The results are shown in Figs.~\ref{visc_2}
and~\ref{visc_1} (C).  The choice of the mean radius at which to
evaluate the spectrum has no qualitative effect on the results.  Both spectrums
show rapid decay for high mode numbers, with more efficient decay for the lower
dynamic Reynolds number set.

Overall, viscous dissipation does not effect the radial, symmetric
implosion until exponentially small length-scales, but it has a dramatic
effect on the vibrational mode behavior.  The chirping characteristic of the
 vibrations persists, but the freezing of the mode amplitudes is lost and
a rapid decay of the mode amplitudes occurs for high $n$ and low
dynamics Reynolds number.


To connect this result back to the experiments, we note that
in the bubble breakup experiments, surface
tension is effective in smoothing out the surface and removing any
higher mode distortions which may be present, because the early dynamics
is completely dominated by surface tension before the implosion
begins~\cite{burton05,keim06,thoroddsen07}. But, if a short
wavelength disturbance was present, viscous dissipation would be very
efficient at wiping it out (Fig.~\ref{visc_1}). So in these experiments
we expect only low mode number distortions to be present.

The solid body impact experiments do not have such a surface
tension dominated regime and thus short wavelength disturbances may be
present in the initial shape.  With the faster radial implosion and
higher dynamics Reynolds number in the experimental regime,
viscous dissipation would cause decay of the highest modes but it is
less effective here than in the bubble experiments so some higher modes
could still be present, which could be part of the reason why
the intricate textures and swarms of satellite bubbles are observed
in those experiments~\cite{bergmann06}.

\section{Rotational motion}
Introducing a small rotational flow into the external liquid provides a
clever way to
remove the singularity formation that occurs as the radius
of the void collapses to zero.
This does not break
the cylindrical symmetry, but causes the liquid to swirl around
as the hole collapses.  A consequence of the rotation is that finite amount
of energy is always stored in the rotation flow because of angular
momentum conservation, and a barrier or turning
point in the radial motion is encountered (like in
 central force motion~\cite{landaumech}) ~\cite{rogers97}.  The turning
point occurs at a non-zero but exponentially small radius (for a weak
rotational flow), introducing an additional length- and time-scale to
the implosion process.
Thus, the rotational flow allows us to examine the consequences
of the singularity formation (and decoupling of the implosion process
from external scales) on the vibrational mode behavior.

For the axisymmetric, imploding cylinder, we now allow the
liquid to be swirling inwards, possessing angular momentum.
The velocity field then
has an additional rotational component which can be written as
$u_{\theta}=\Gamma/2 \pi r$.
Kelvin's circulation theorem states that the circulation
$\Gamma$ around a closed loop for potential flow is a constant,
\beq
\Gamma=\oint \bu\cdot\mathbf{dl},
\eeq
determined by the initial conditions.  Thus, the dynamics remains
integrable with an additional conserved quantity, the circulation.
The Hamiltonian in (\ref{ham}) must include
the additional kinetic energy from the rotational flow. This
energy is easily found by integrating $u_{\theta}$ over the
liquid region and the total kinetic energy is now
\beq
E=\frac{ M(R)}{2} \dot{R}^2+\frac{M(R)}{2} \left(\frac{\Gamma}{2\pi R}\right)^2
\label{ecirc}
\eeq
The evolution equation for $R(t)$ (\ref{RP}) is modified
with an additional term on the right hand side, $\rho \Gamma^2/8\pi^2R^2$.

The consequence of the rotation is that
a singularity in the governing equations no longer occurs.
Instead, a turning point in the motion occurs when $\dot{R}=0$
which can be quickly seen from (\ref{ecirc}) to be exponentially
small unless the rotational flow is large:
$R_{turn}/R_{\infty}=\exp(-4\pi E/\rho \Gamma^2)$,
where we have used the form for the mass from
Chapter~\ref{axi}, $M(R)=2 \pi \rho R^2 \ln(R_{\infty}/R)$.
This change occurs because the rotational flow contains a finite,
and growing amount of energy as the hole collapses, preventing
the radial velocity from diverging.

Upon again performing the linear stability analysis for the azimuthal shape
perturbations, which now includes terms depend on the rotational flow $u_{\theta}$,
we find a description in terms of only the $\cos(n\theta)$ harmonics is
insufficient to describe the dynamics.  We must now include the second
linearly independent $\sin(n\theta)$ harmonics and write the perturbed interface
as:
$S(\theta,t)=R(t)+\sum_n(b_n(t)\cos(n\theta)+c_n(t)\sin(n\theta))$.
The vibrational modes $b_n$ and $c_n$ are coupled by their evolution
equations:
\beq
\ddot{b}_n +\left(\frac{2\dot{R}}{R}\right)\dot{b}_n+(1-n)\left( \frac{\ddot{R}}{R}
+n\left(\frac{\Gamma}{2\pi R^2}\right)^{2}\right)b_n+2n\left(\frac{\Gamma}{2\pi R^{2}}\right)\dot{c}_n=0,  \label{bn}
\eeq
\beq
\ddot{c}_n +\left(\frac{2\dot{R}}{R}\right)\dot{c}_n+(1-n)\left( \frac{\ddot{R}}{R}
+n\left(\frac{\Gamma}{2\pi R^2}\right)^{2}\right)c_n-2n\left(\frac{\Gamma}{2\pi R^{2}}\right)\dot{b}_n=0. \label{cnn}
\eeq

\begin{figure}
  \psfrag{aaa2}{$b_{10}/b_0, c_{10}/c_0$}
  \psfrag{RRRR}{$R/R_0$}
\begin{center}
\includegraphics[width=0.7\textwidth, angle=0]{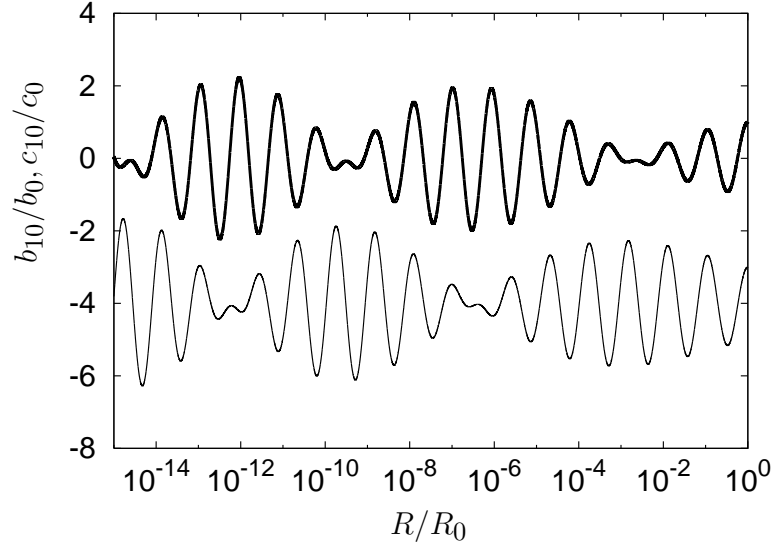} 
\end{center}
\caption{Coupled mode amplitudes with presence of swirl flow for $n=10$,
  $b_{10}$ (thick line) and $c_{10}$ (thin line, displaced downwards for clarity). The
  initial conditions at $R=R_0$ are $b_{10}=c_{10}=b_0$ with zero derivative.
  The relative swirl flow strength is $u_{\theta}/u_r\approx 10^{-2}$ (over the range
  of the plot it increases only to $3\cdot 10^{-2}$ as $R$ decreases). $R_{\infty}/R_0=10^2$
and the implosion energy is about }

\label{swirl}
\end{figure}

For a weak ($\rho \Gamma^2/E\ll1$) rotational flow the turning point
occurs at an exponentially small radius, so the leading order radial implosion dynamics
are unaffected until $R=O(R_{turn})$, outside the limits of a continuum theory.  However, the
vibrational mode behavior does experience a fundamental change -- rather
than forming standing waves with shapes that pulse as $R\rightarrow 0$, the rotational
flow induces traveling waves around the circle, again with frequencies proportional to
$1/R$.  Numerical results for $b_n$ and $c_n$ for $n=10$ are shown in Fig.~\ref{swirl}.
The fast vibration is the original chirping described in Chapter~\ref{aziperts} with
frequencies $\sqrt{n-1}/R$, and the slower oscillation in the amplitude envelope
is actually caused by the spinning of the quickly pulsing shapes as the hole closes.

To simply see this behavior in (\ref{bn}) and (\ref{cnn}),  assume the radial collapse
dynamics follow $R=\alpha \tau^{1/2}$ and  $u_{\theta}/u_{r}=\Gamma/\pi \alpha^2\ll 1$. Then
trying \textit{a priori} solutions of the form
$b_n=b_0 \cos(\omega_n \ln(R))\cos(\sqrt{n-1}\ln(R))$ and
$c_n=b_0 \sin(\omega_n \ln(R))\cos(\sqrt{n-1}\ln(R))$ yields
$\omega_n=\frac{n \Gamma}{\pi \alpha^{2}}$. The full shape can then be written
(without the phases)
\beq
S(t,\theta)=R(t)+\sum_n b_{n,0}\cos[\omega_n \ln(R)-n\theta]\cos(\sqrt{n-1}\ln(R)),
\eeq
clearly showing the traveling wave nature of the solutions, with $n$-independent
frequencies $(\omega_n/n)R^{-1}=(\Gamma/\pi \alpha^{2})R^{-1}$.  This is the
frequency of the slow oscillations of the
envelope of the the faster vibrations in Fig.~\ref{swirl}.
Even though the ratio $\Gamma/\pi \alpha^{2}$ is not perfectly constant as was assumed
in the simple scenario above, but involves logarithmic corrections,
 it changes so slowly that it increased by only a factor
of 3, over the 10 decades collapse of $R$.  This is reflected in the visible
weak variation of the long frequency.

Introduction of the weak rotational flow did not destroy the vibrational mode
amplitude freezing behavior but did cause the emergence of another frequency
in the dynamics, the $n$-independent frequency of the spinning motion of the
shapes.  The surface spins faster and faster as void closes, while the shapes
vibrate with constant amplitude.  The above analysis assumes
the rotation is slower than the radial converging flow, and thus the effects of the
finite turning radius turned out to be negligible.  It would be interesting to fully examine
the asymptotic behavior when the rotational flow becomes the same size as
the radial flow, and then the destruction of the singularity formation would likely
cause rapid amplitude growth or decay.

In the bubble experiments which start with a quiescent liquid bath, there is likely
very little circulation present, though it is not possible to completely eliminate.
In this case, the presence of a weak and unintentional rotational flow are not
damaging to the vibrational mode behavior, but simply cause a slow spinning
of the vibrating shapes.
Such a rotational flow could be the cause behind observations of rotating surfaces and
satellite bubbles~\cite{keim, thoroddsen08}.

\section{Surface tension}
\begin{figure}
\subfigure[][\ $n=10$]{
  \psfrag{aaaa}{\large{$b_{10}/b_0$}}
  \psfrag{RRRR}{$R/R_0$}
  \includegraphics[trim=0 0 0 2mm, clip, width=0.5\textwidth, angle=0]{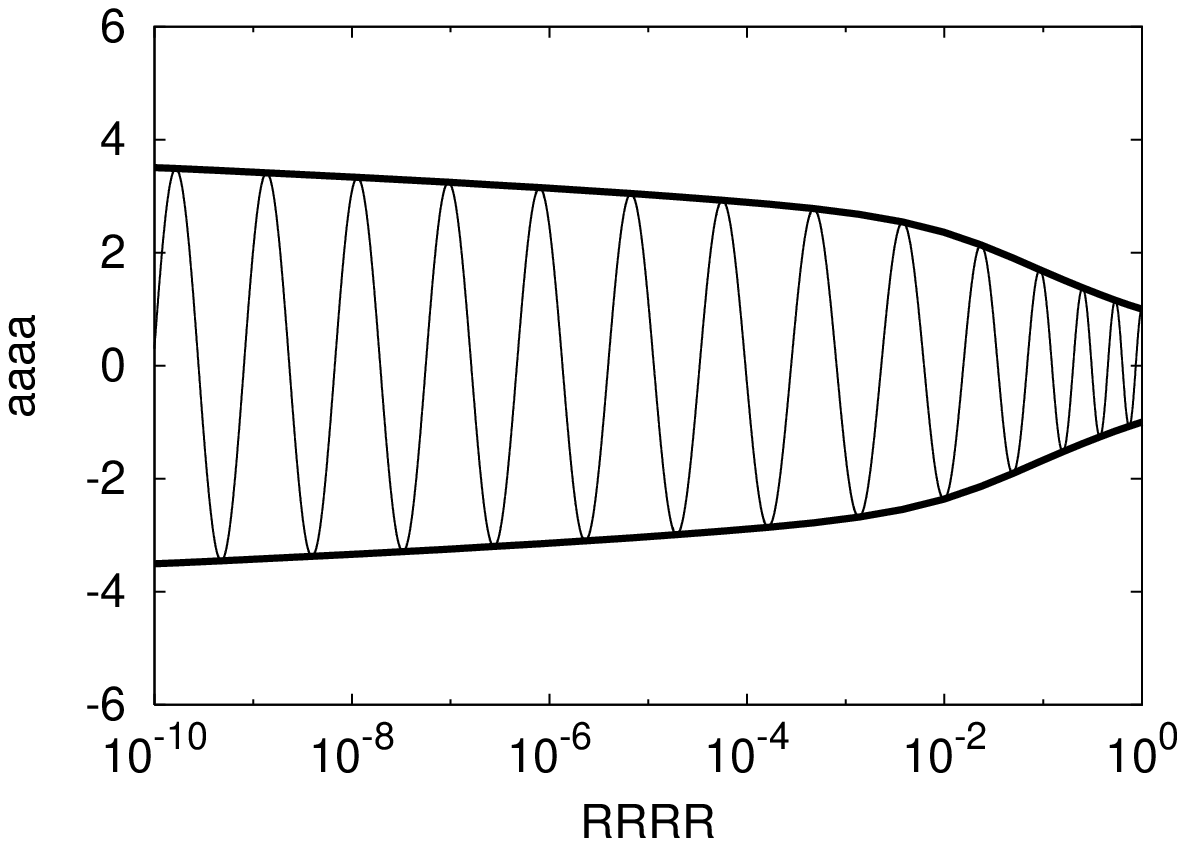}}
\subfigure[][\ $n=2-30$]{
  \psfrag{aaaa}{\large{$b_n/b_0$}}
  \psfrag{RRRR}{$R/R_0$}
  \includegraphics[trim=0 0 0 2mm, clip, width=0.5\textwidth, angle=0]{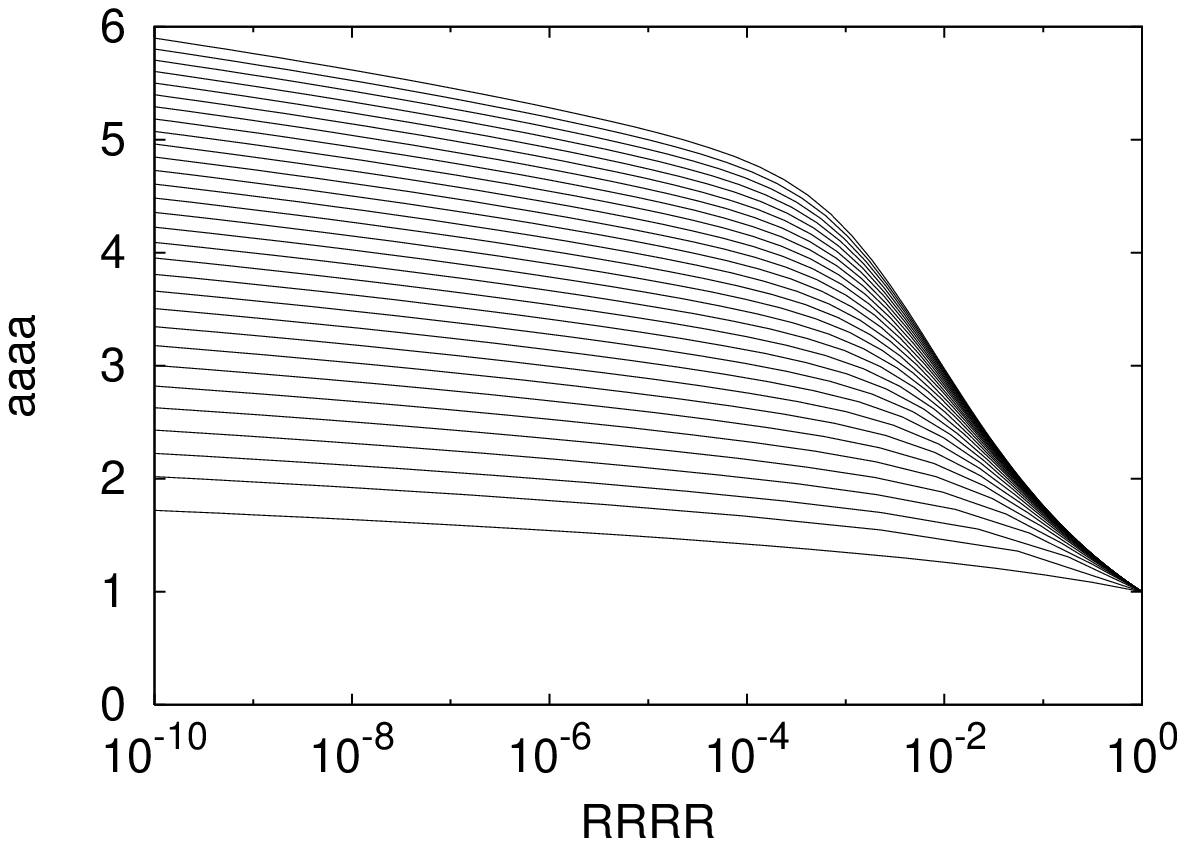}}
\subfigure[][\ spectrum]{
  \psfrag{aaaa}{\large{$b_n/b_0$}}
  \psfrag{n}{$n$}
  \psfrag{xxxx2}[cc][cl]{\footnotesize{$R/R_0=10^{-2}$}}
  \psfrag{xxxx5}[cc][cl]{\footnotesize{$R/R_0=10^{-5}$}}
  \includegraphics[trim=0 0 0 2mm, clip, width=0.5\textwidth, angle=0]{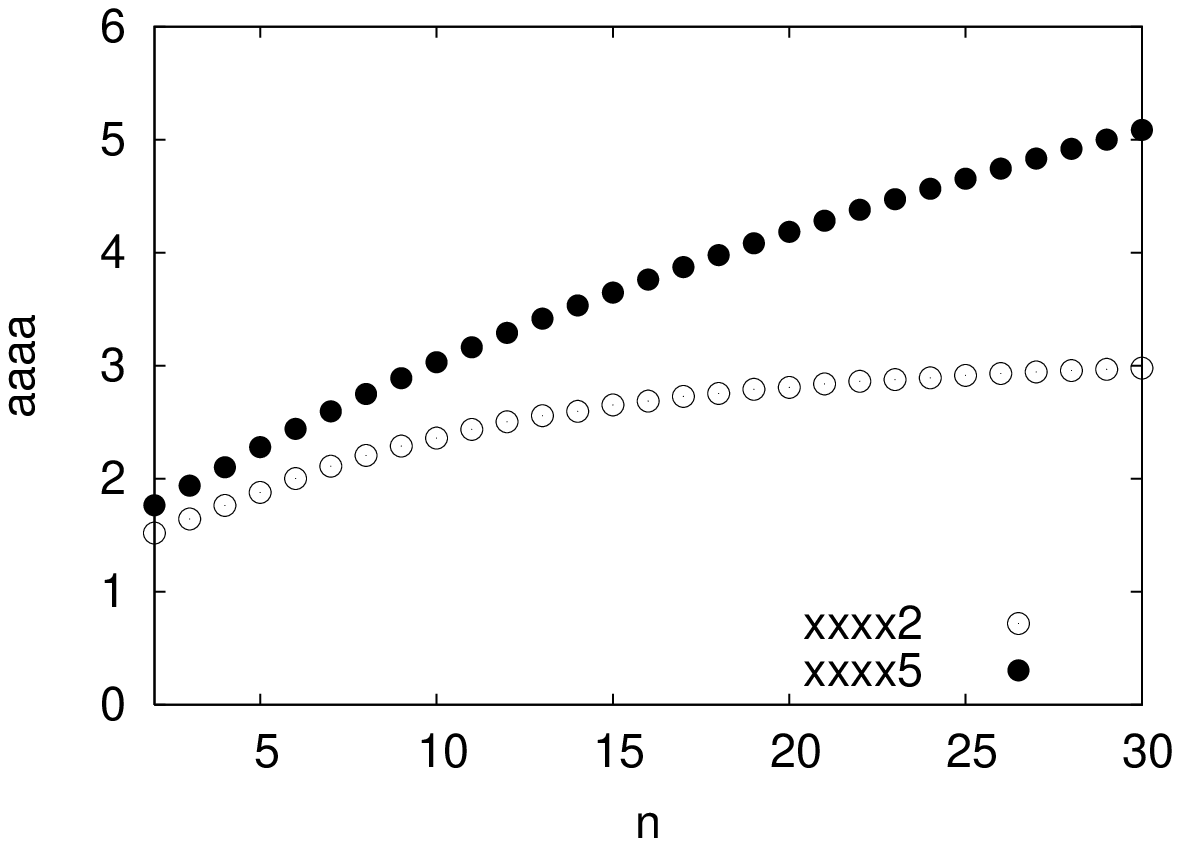}}
\caption{Effect of surface tension on early (large $R$)
vibrational mode behavior. Here, $R_{\infty}/R_0=10$ and $E/\pi \gamma R_0=10$.
The amplitudes are rescaled by their initial values $b_0$ and the mean radius
is rescaled by its initial value $R_0$.
(a) Representative vibration for $n=10$ (thin line) shows that surface tension
 causes an initial increase in the amplitude (envelope is thick line)
and frequency of vibration.
After the mean radius has decreased beyond $R_{10}^{crit}\approx R_0/10^2$, the
vibration settles in to the asymptotic inertial state with amplitudes
and periods as in Fig.~\ref{inertialmodes}.
(b) Envelope of vibrational mode amplitudes for $n=2-30$ (lower- to upper-most).
(c) Amplitude spectrum taken at $R/R_0=10^{-2}$ (open circles)
    and at $R/R_0=10^{-5}$ (solid circles).}
\label{stmodes}
\end{figure}

Surface tension forces become negligible to the radial implosion in the
asymptotic inertially dominated
regime as seen in (\ref{ham}) and verified by experimental measurements of the
implosion dynamics.  However, here we will see that the effect of surface tension on the
inertial vibrations within the experimental regime, is to cause a short-lived but
rapid growth in mode amplitude and increase in frequency in the experimental
regime before converging onto the asymptotic form given in (\ref{agov}).
We modify (\ref{agov})
so that the local curvature is no longer simply $1/R$ but to first
order in $b_n/R$: $\kappa\approx\frac{1}{R}+(n^{2}-1)b_n \cos(n\theta)/R^{2}$,
yielding
\beq
\ddot{b}_n+\left(\frac{2\dot{R}}{R}\right)\dot{b}_n
          +\left(\frac{\ddot{R}}{R}(1-n)
                 +\frac{\gamma n (n^{2}-1)}{\rho R^{3}}\right)b_n=0.\label{ast}
\eeq
This adjusts the local Laplace pressure jump as we go around the shape through
the peaks and valleys.
When the radial dynamics is inertially dominated, the $b_n$ coefficient
in (\ref{agrowth}) now includes $\gamma n(n+1)\pi R \ln(R_{\infty}/R)/E$.
By comparison of the original inertial term in this coefficient,
we see that the cutoff for when
surface tension will no longer affect the vibrations is when $R<R_n^{crit}$ with
the critical radius
\beq
R_n^{crit}\ln(R_{\infty}/R_n^{crit})=\frac{E}{\gamma \pi n (n+1)}.
\eeq
We numerically integrate the solutions for $b_n(t)$ using the radial
dynamics from a numerical solution of (\ref{energybalance}). The solutions show that the effect
of surface tension is to speed up the vibrations and give additional short-lived
amplitude growth when the mean radius is large (see Fig.~\ref{stmodes}). Fig.~\ref{stmodes} (A)
shows the growth and frequency increase for an $n=10$ vibration.  The amplitude
envelopes for $n=2-30$ are plotted in Fig.~\ref{stmodes} (B), and the spectrum (using the
same definition for the case of viscous dissipation above) in (C).
Higher modes are impacted more severely and for longer into the implosion,
since the critical radius $R_{crit}$ decreases with increasing $n$.

The bubble pinch-off experiments are unique in the fact that they
are initially dominated by surface tension (which will efficiently smooth
out high modes), but by the time the mean radius
reaches $O(300 \mu$m) the radial implosion is inertial.  This gives surface
tension a short time to act on the vibrational modes before the inertial
limit is reached for $R<R_n^{crit}$.


\chapter{Connection between theory and experiments}
The preceding chapters laid out a scenario of how and why a nearly
cylindrical void collapsing in an inviscid liquid retains an imprint
of initial distortions to its shape.  The cylindrically-symmetric
model of Chapter~\ref{axi} predicted how the void radius should
collapse to a point, and the stability analysis of Chapter~\ref{aziperts}
predicted that and initial distortion causes vibrations of
the shape as the implosion proceeds.
The vibration amplitudes of all wavelength
disturbances become frozen, and eventually overtake the leading-order
symmetric implosion, generating strong distortions.
Consideration of viscous
dissipation, rotational flow, and surface tension effects in Chapter~\ref{phys}
predicts that the mode
amplitude freezing and chirping (frequencies $\propto 1/\tau$)
are characteristics within the realm of experimental observation.

Below we briefly review two types of experiments that generate
void collapse in water.  Then, because
the vibrational behavior is inherently linked to the radial implosion
dynamics via (\ref{agov}) we show the quasi-2D is capable of
describing the evolution of 3D experimental profiles. Finally,
we directly compare an experimentally observed shape vibration
to the prediction of (\ref{agov}) for an $n=2$ vibrational mode.

\section{Review of experiments}
In the bubble release experiments a bubble is grown quasi-statically by
slowly pumping gas through a nozzle~\cite{burton05,keim06,thoroddsen07}.
Surface tension keeps the bubbles
connected to the nozzle until it can no longer stabilize the surface
against the upwards pull of buoyancy.  At this point, the neck of the bubble
rapidly pinches off.  The entire disconnection process occurs on a ms time
scale, much faster than the rising motion of the bubble.  The earliest
dynamics are determined by a competition between
surface tension and gravity. However, once the implosion begins, the
inertia of the water rushing in to fill the void quickly starts to
dominate the process, evidenced by the time evolution of the minimum
radius $h_{min}$ following close to a 1/2 power law in time, rather
than $2/3$ if surface tension were to drive the flow.

For the disc impact experiments, a solid disc is rapidly drawn into a bath
of water, creating a cavity in its
path~\cite{bergmann06, duclaux07, bergmann08}.  Because of the high impact
velocities used, the initial Weber number, which characterizes inertial effects
relative to surface tension is greater than $10^2$ ($We=\rho U^2 h/\gamma$;
$\rho$ is the liquid density, $\gamma$ is the surface tension of the
gas/liquid interface, $U$ a characteristic velocity scale,
$h$ a characteristic length scale).  Thus there is no regime here where
surface tension can efficiently act to smooth out distortions to the
cavity shape.

Both types of experiments observe a medley of different kinds of
disconnection events depending on the specifics of the set-up, geometry and
orientation of the nozzle or disc, and control parameters.
In the bubble experiments, side-by-side satellite
bubbles remain after disconnection if the nozzle is not carefully
leveled~\cite{burton05, keim06}. Fig. ~\ref{bubbles} (C)
shows the bubble neck before and after disconnection when the nozzle
has been purposely tilted by $2^{\circ}$. Even more exotic
and highly asymmetric shapes are observed if the bubble is not grown
quasi-statically from the circular nozzle, but is instead blown
rapidly through a slot shaped nozzle~\cite{keim06}.
Often times in the impact experiments, the surface before
disconnection appears textured, and after disconnection leaves
behind a disordered swarm of satellite bubbles,
rather than cleanly disconnecting at a point~\cite{bergmann06, duclaux07}.
These observations signal an instability in the underlying
dynamics to azimuthal and/or axial perturbations, which
prevents the axisymmetric implosion from being realized under
generic conditions.
Similar effects have not been observed in experiments where
surface tension is the force driving breakup.
For example, a water drop breaking in air
appears to always break in a universal way, independent of
the exact experimental set-up~\cite{shi94}.
Despite the clear evidence for the initial distortions in the
azimuthal direction playing a critical role in the dynamics
before and after bubble disconnection, it remained unexplored in the
context of cylindrical voids until recently~\cite{schmidt08}.
The accumulation of experimental results suggests that the
fixed point for axisymmetric cavity collapse proposed~\cite{egg07}
is not stable once the azimuthal direction is considered.

The final stages of both implosion situations have the same inertia-
dominated dynamics, but the initial states are quite different.
The impact experiments never pass through a surface tension dominated
regime, whereas the bubble experiments always start there.  This
puts both experiments in unique positions.  For impact,
large amplitude and/or short wavelength
distortions can be applied by using different shaped discs,
because surface tension will not smooth them out.  However, the advantage
in the bubble experiments is the possibility to generate smooth,
low wavelength, single mode perturbations to the shape.


\section{Axisymmetric implosions}
\begin{figure}
\subfigure[][]{
  \label{hmin}
  \psfrag{hminxxxx}{\large{$h_{min}(\mu m)$}}
  \psfrag{timexxxx}{time $(\mu s)$}
  \includegraphics[trim=0 0 0 1mm, clip, width=0.7\textwidth]{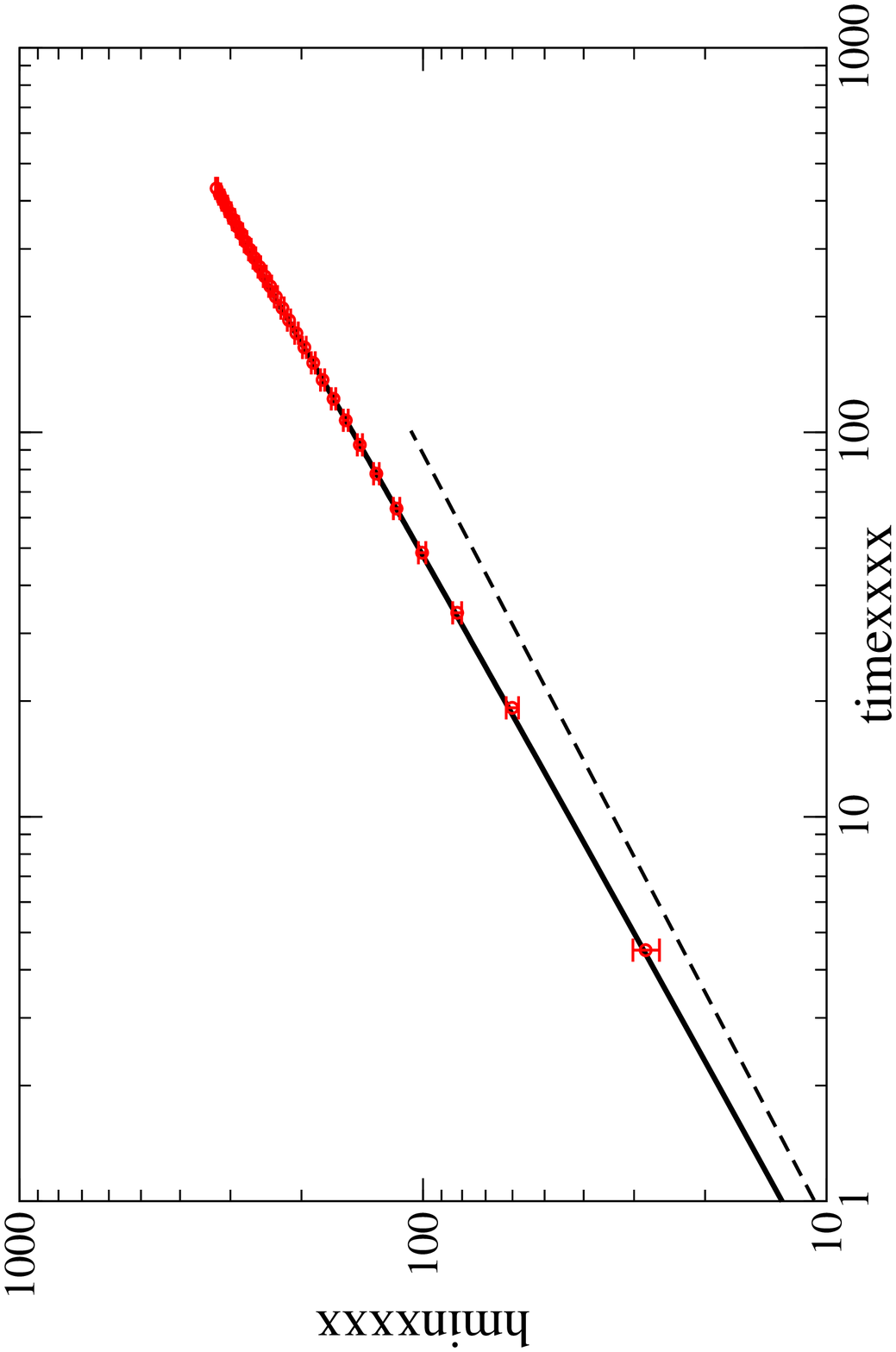}}
\subfigure[][]{
  \label{profiles}
  \includegraphics[width=0.7\textwidth, angle=0]{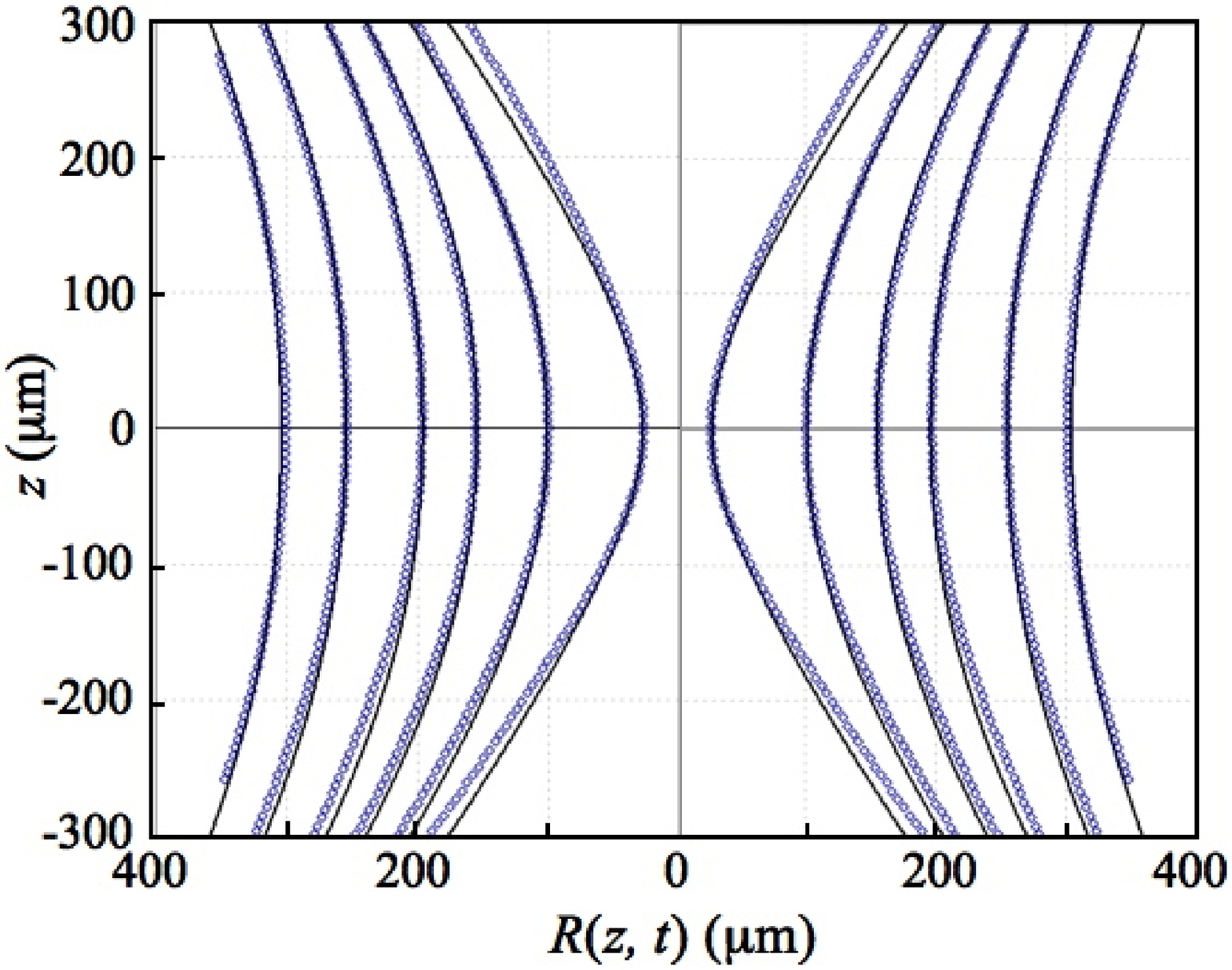}}
\caption{(a) Collapse of minimum radius for air bubble released from circular,
4mm diameter nozzle (red circles).  Theoretical inertial implosion from
(\ref{energybalance}) with an implosion energy per unit length of $E=12$ dyne
(black line).  Dashed line shows slope of $1/2$.
(b) Experimental neck profiles of the same air bubble (blue diamonds).
Theoretical profiles calculated as described in text using conservation
of energy for each isolated layer (lines). Profiles are at 4.5, 49, 108,
166, 270, and 372 $\mu$s before disconnection (inner to outer curves).
Experimental data provided by Nathan Keim (U. of Chicago).}
\label{profile4mm}
\end{figure}

With the implosion dynamics outlined above we can now compare the profiles
predicted by conservation of energy to those actually observed in experiments.
Neck profiles from traces of experimental images of a bubble released from a 4 mm
diameter circular nozzle are shown in Fig.~\ref{profile4mm}~(b).
To generate the theoretical profiles for comparison we first use the time-dependence
of the minimum (at $z=0$)  to define the implosion energy per unit
length.  Within the range shown, the radius is
following close to a $1/2$ power law, signaling that the radial
implosion is now far within the inertially dominated regime.  We then
compare the evolution of the minimum to that predicted in the inertial
regime, from numerical integration of
(\ref{energybalance}) to determine the implosion energy $E$.

Taking $R_{\infty}=2$mm, a physically
relevant length scale as it is the radius of the nozzle,
we find $E=12$ dyne gives the best fit
to the minimum (Fig. ~\ref{profile4mm}~(a)).  The result is
insensitive to the value of $R_{\infty}$ used and good fits within
experimental error can be found for $R_{\infty}=1$mm with $E=8.7$ dyne
or up to $R_{\infty}=4$mm with $E=15$ dyne.  These results would be
indistinguishable by eye compared to those shown in Fig. ~\ref{profile4mm}~(a).

The implosion energy of $12$ dyne was then applied at every height to track the
full profile evolution.  Because of the time invariance in the problem,
the symmetric form of the profile closest to disconnection
(innermost profile, averaged about $z=0$) was used as the initial state,
and then the expansion of the radius at each height was evolved according
to $\dot{R}=+\sqrt{E/\pi \rho R^2 \ln(R_{\infty}/R)}$.
Though the angles observed in bubble experiments typically are not small
(half angles range from about 25-35 degrees depending on conditions),
the quasi-2D model does a surprisingly good job describing the complete 3D
surface evolution. This suggests the errors involved in applying the
quasi-2D model are quite small over the finite range in $h_{min}$ afforded
by the experiment, though this remains to be checked via numerical simulation.

One aspect of the experiments which is quite obviously
not captured in the model is the development of an up/down
asymmetry in the cones above and below the minimum.
The upper cone has an opening angle which is larger than the lower by
about 5 degrees, but the precise amount depends on the vertical range
used to define the angle.
One might think that this could be simply accounted for within
the framework of the 2D model by incorporation of a hydrostatic
pressure gradient, which would break the axial symmetry about the
minimum.

To introduce the hydrostatic
pressure gradient $-\Delta \rho g$ we simply include it in the
Hamiltonian (\ref{ham}) so the pressure term becomes
$(\Delta P -\Delta\rho g z)\pi R^2$. Then,
the energy is a function of $z$, with deeper layers having larger potential
energies because the hydrostatic pressure at $r=R_{\infty}$ is higher there.
If the radius at the same height above and below $z=0$ are equal, the
one with higher energy has a faster implosion rate (\ref{energybalance}). Thus,
na\"{\i}vely, one might expect that the cone below the origin will implode
faster and have a smaller angle relative to the upper cone.
However, one finds that this is not the case.  Instead of causing a change in the
cone angles, the primary effect is a downwards linear translation of the profile
with the angles unaffected.

The profiles can be shifted to follow this translation
of the minimum and are of the form $R(\psi)^2=h_{min}^2+2 c_0 H_0 \psi^2+ c_2 \psi^4$
where $h_{min}$ is the radius of the minimum at $z=z_{min}$ and $\psi=z-z_{min}$.
The cone angles remain $\sqrt{2 c_0 H_0}$.  The only possibility for the observed
difference in apparent cone angles is if the $\psi^3$ term is very large, which requires
the hydrostatic energy to be a substantial component of the implosion energy.  This
case is relevant to the larger-scale impact experiments~\cite{duclaux07,bergmann06} but
not for bubble pinch-off where the scale of the neck is $O(100\mu$m).
An additional inconsistency is fact that if the difference in cone angles
was caused by a large hydrostatic pressure gradient, the motion of the minimum would be
downwards, but the experiments observe and the upwards motion
of the minimum. Within
the framework above, the hydrostatic pressure alone is not sufficient to
explain the difference in cone angles and upwards motion of the minimum
observed in the experiments.
The up/down asymmetry which appears may be partly due to the slow upwards motion
of the buoyant bubble, but is likely related to non-local flow coupling or energy
transfer in the vertical direction, and is a point to be examined further in the future.


\section{Predicted and observed vibrational mode}
Recent experiments at the University of Chicago
have been able to observe vibrations of the bubble neck shape
as it implodes by releasing a bubble from an oblong nozzle~\cite{schmidt08, keim}.
As the bubble grows from the nozzle, the neck is initially elongated in the
same direction as the slot. Though surface tension is efficient in rounding up the
bubble neck, it does not completely erase the effect of the highly asymmetric nozzle,
and a small asymmetry in the cross section can be seen.

\begin{figure}
\psfrag{Rbarxx}{$\bar{R}(\mu m)$}
\psfrag{Rbarxxx}{$\bar{R}(\mu m)$}
\psfrag{tauxxxx}{$\tau(\mu s)$}
\psfrag{anxxxx}{$a_2(\mu m)$}
\begin{center}
\includegraphics[width=0.85\textwidth, angle=0]{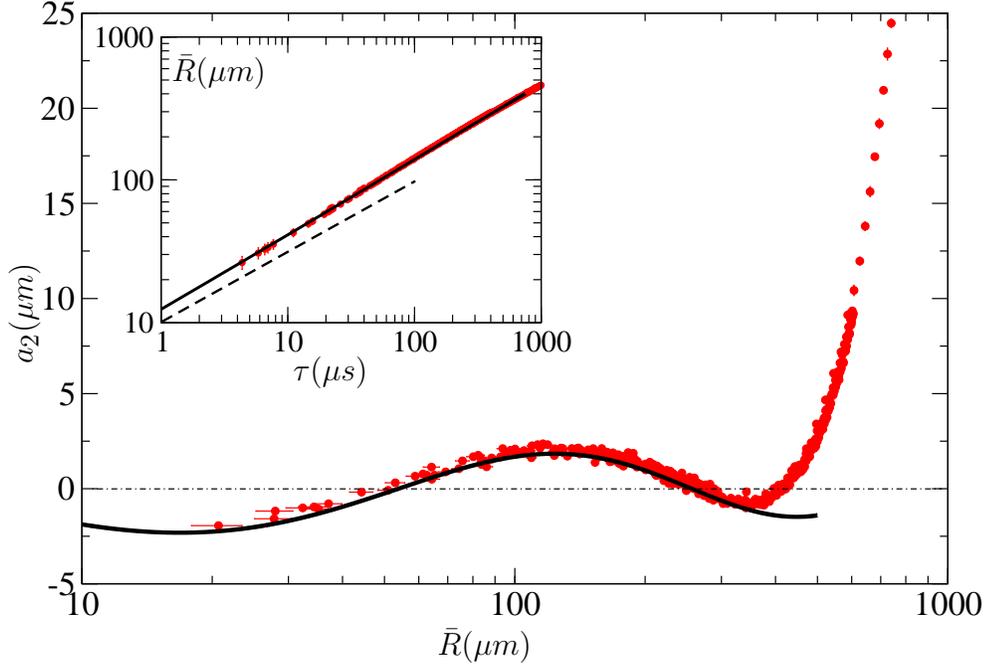}
\end{center}
\caption{Experimentally observed vibrations of $n=2$ mode by release
from 1.6 mm by 9.6 mm slot nozzle (red).
Theoretical vibration calculated from (\ref{ast}) as
described in text (solid black line).  Inset: Experimental mean radius collapse (red)
corresponds to an inertial implosion with $E=12$ dyne, and $R_{\infty}=5mm$ (solid black).
A power-law of 1/2 is shown by the dashed line for comparison.}
\label{vibrations}
\end{figure}

As the neck begins to rapidly implode, the neck shape  vibrates so
that at a later time it will be elongated in a direction opposite to slot.
Depending on the speed at which the bubble is released from the slot nozzle
(either quasi-statically or via a rapid burst of air) different numbers of periods
can be observed.  Fig.~\ref{vibrations} shows these vibrations for quasi-static
release from a slot of size 1.6 mm by 9.6 mm.  By recording the implosion with
two high speed cameras at a 90 degree angle, the perturbation amplitude
is simply the half the difference in the measured widths from the two views
$a_2=(h_{side}-h_{front})/2$, and the mean radius is $\bar{R}=(h_{side}+h_{front})/2$.

Above a measured mean radius of $\bar{R}\approx 400\mu$m
there is a rapid decay of the distortion amplitude which is likely due to the
fact that the dynamics is truly three-dimensional and dominated by surface tension in
this regime.  Around $\bar{R}\approx 300 \mu$m a vibration centered around
$a_2=0$ sets in and continues until the last possible data point when
$\bar{R}\approx 20\mu$m. Below $\bar{R}\approx 10\mu$m there is evidence of some
weak decay in the amplitude, which could be due to viscous dissipation.
Though the vibration shown in Fig.~\ref{vibrations} is at height of the minimum
radius in the experiment, it is also possible to observe the vibrations
vibrations at different heights along $z$.

Below we outline the parameters, which are consistent with the
experiment, used in the model to generate the theoretical vibration shown in
Fig.~\ref{vibrations}.  We begin by defining the total energy of the symmetric
radial collapse which reproduces the radial inertial implosion
dynamics for $\bar{R}(t)$.   For the data shown in the inset to Fig.~\ref{vibrations},
this value is $E=12$ dyne/cm when $R_{\infty}=5$mm, the average radius of the slot.
Similarly to the data in Fig.~\ref{profile4mm}, while $R_{\infty}$ is chosen
within a physically meaningful range as the system size, the radial dynamics
from energy conservation (\ref{energybalance}) describe the observed trajectory.
Below we estimate the appropriate value for $\gamma$ to be used in the
evolution equation for $a_2(t)$ (\ref{ast}) which will take into account
the axial curvature contribution to the Laplace pressure jump across the interface as
done in (\ref{laplace}).

From examination of experimental images of the data set used to generate
the data shown in Fig.~\ref{vibrations} we find that $R_{ax}/h_{min}\approx 2.8$.
The value of 2.8 was chosen by taking the average of the ratio  $R_{ax}/h_{min}$
over the range of the experimentally observed vibration. Thus, to effectively include the
axial contribution, a value of $\gamma=(1-1/2.8)\gamma_0=0.64\gamma_0=46$ dyne/cm
where $\gamma_0$ is the surface tension of an air water interface at room temperature,
$72$ dyne/cm, was used in the numerical solution of (\ref{ast}).
The black line plotted in Fig.~\ref{vibrations} is the result of a numerical
solution to (\ref{ast}) with the parameters outlined above.
The amplitude and phase (not predictable in this model) were chosen to fit
the peak around $\bar{R}=100\mu$m.



\chapter{Conclusion}
The 2D model of the implosion of a cylindrical void in water is an integrable
Hamiltonian system with special properties.
These properties include conservation
of the implosion energy and an exceptional perturbation mode spectrum which
freezes as the implosion proceeds, and causes constant amplitude inertial
vibrations about the radial implosion.
This ideal system is closely related to the dynamics of the thin neck of air which collapses
as an air bubble disconnects from an underwater nozzle.  The memory
mechanism in the simplified model is capable of explaining why asymmetries are
observed to persist in experiments, and has predictive power for how
azimuthal disturbances on the surface can dramatically alter the final stage of
the pinch-off process.

The main features (amplitude freezing and vibrational chirping)
of the inertial vibrations are robust against
additional effects of surface tension, viscous dissipation, and rotational flow
within the experimental regime, and thus we are able to directly
compare the predicted vibrational mode to one excited in experiments.
However, when comparing the observed amplitude and frequency of the $n=2$
vibration, the most important effect was found to
be surface tension. Inclusion of surface tension effects via the Laplace
pressure jump across the interface acted to speed up the vibrations for
a short time within the experimental regime.

At this point, a variety of physical effects were considered separately, to
isolate the individual consequences on the vibrational mode spectrum. There is
the potential for interesting mode spectrums to arise if multiple effects are
combined.  For example, the form of the normal stress balance on the
interface will change substantially if both viscous dissipation and a rotational
flow are included.  It may also be possible to find peaks in the spectrum, where
one wavelength grows more rapidly than others, when combining surface tension
(which causes amplitude growth) with viscous dissipation (which causes amplitude
decay). More exploration of the outcome of these possibilities will be necessary
to predict vibrational mode behavior for a larger range of experimental
parameters.




\newpage
\addcontentsline{toc}{chapter}{References}
\begin{singlespace}
\bibliography{bubblecite}
\bibliographystyle{naturemag}
\end{singlespace}


\end{document}